\documentclass[5p]{elsarticle}

\usepackage{lineno,hyperref}
\usepackage{dblfloatfix} 
\usepackage{float}
\usepackage{graphicx}
\usepackage{soul} 
\usepackage{xcolor} 
\usepackage{caption}
\usepackage{amssymb}
\usepackage{amsmath}
\usepackage{bm}
\usepackage[normalem]{ulem}
\usepackage{upgreek}
\usepackage[detect-all]{siunitx}
\journal{ArXiv}

\begin{document}


\begin{frontmatter}

\title{Micromechanics reveal strain rate dependent transition between dislocation mechanisms in a dual phase high entropy alloy}


\author[myaddress2,mymainaddress]{Szilvia Kal\'{a}cska\corref{mycorrespondingauthor2}}\cortext[mycorrespondingauthor2]{Corresponding author} \ead{szilvia.kalacska@cnrs.fr}

\author[mymainaddress]{Amit Sharma}
\author[mymainaddress,address3]{Rajaprakash Ramachandramoorthy}
\author[address4]{Ádám Vida}
\author[address5]{Florian Tropper}
\author[address5]{Renato Pero}
\author[address5]{Damian Frey}
\author[mymainaddress]{Xavier Maeder}
\author[mymainaddress]{Johann Michler}
\author[address6]{Péter Dusán Ispánovity}
\author[myaddress2]{Guillaume Kermouche}

\address[myaddress2]{Mines Saint-Etienne, Univ Lyon, CNRS, UMR 5307 LGF, Centre SMS, 158 cours Fauriel 42023 Saint-Étienne, France}

\address[mymainaddress]{Empa, Swiss Federal Laboratories for Materials Science and Technology, Laboratory of Mechanics of Materials and Nanostructures, CH-3602 Thun, Feuerwerkerstrasse 39. Switzerland}

\address[address3]{Max-Planck-Institute für Eisenforschung, Max Planck Strasse 1, 40472 Düsseldorf, Germany}
\address[address4]{Bay Zoltán Nonprofit Ltd. for Applied Research, Kondorfa u.1., H-1116 Budapest, Hungary}
\address[address5]{Alemnis AG, Business Park Gwatt, Schorenstrasse 39, 3645 Gwatt (Thun), Switzerland}
\address[address6]{ELTE E\"{o}tv\"{o}s Lor\'{a}nd University, Department of Materials Physics, P\'{a}zm\'{a}ny P\'{e}ter s\'{e}tany 1/a, 1117 Budapest, Hungary}

\begin{abstract}
An equimolar NiCoFeCrGa high entropy alloy having dual-phase homogeneous components was studied, where the constituent phases exhibit distinct mechanical properties. Micropillars with various diameters were created from two differently heat treated samples, then they were compressed at slow strain rates, that revealed the material's limited sensitivity to size. On the other hand, increased strain rate sensitivity at high deformation speeds was observed, that differs substantially depending on the phase composition of the specimen. Dislocations within the two phases were studied by high resolution transmission electron microscopy and high angular resolution electron backscatter diffraction. The performed chemical analysis confirmed that slow cooling during casting create Cr-rich precipitates, that have significant impact on the global strength of the material.

\end{abstract}


\end{frontmatter}


\section{Introduction}

High entropy alloys (HEAs) are multi-principal element intermetallic alloys with at least four constituents with a crystalline structure that lacks long-range ordering. Since their first syntheses \cite{Yeh.2004, Cantor.2004}, they have created significant interest within the materials science community due to their outstanding properties. First and foremost, they provide an exceptional trade-off in terms of high strength coupled with considerable ductility \cite{miracle2019high, george2020high}, but they also exhibit remarkable chemical (such as enhanced corrosion resistance) and thermal properties \cite{tsai2014high, Miracle.2017a} as well as high irradiation resistance \cite{el-atwani.2019}. In addition, due to the large number of constituent elements, the tunability of HEA materials is particularly high \cite{ma2019tailoring, feng2021high}.

As a reult of the large number of elements, their mixing entropy is high, therefore HEAs are mostly created as thermally stable single-phase solid solutions. Still, some HEAs are different in this sense and can exhibit multiple phases in equilibrium \cite{Miracle.2017b, luan2020phase}, that is now making several ways toward \emph{property} and \emph{phase engineering} \cite{chang2020phase, GOU2023118781}. Alloys mainly composed of refractory elements usually have a BCC structure and exhibit high thermal stability and hardness, but are often rather brittle \cite{senkov2010refractory}. On the other hand, HEAs containing mostly $3d$ transition metals in near-equiatomic concentrations usually form an FCC structure, with reduced strength but high ductilty \cite{Miracle.2017b}. In addition, these materials often show serrated flow, that is the sudden and unpredictable drops appearing on the stress-strain curves of random size \cite{chen2018nanoscale}. The combination of two phases in one material may allow one to achieve the best strength-ductility synergy, this is exactly the objective of the present research.

In this paper we focus on a four-component steel-like equimolar NiCoFeCr alloy doped with an $sp$ element (Ga) to make a NiCoFeCrGa HEA. In the literature, the fifth component is most often Al and Cu, but more specialised elements such as Sn, B, Ga, \emph{etc.}\ also occur \cite{Miracle.2017b, Chen.2022}. The speciality of these systems is that while the parent alloy (i.e., the 4-component NiCoFeCr) consists of pure FCC phase with good thermal stability, the doped systems are mostly dual-phase mixtures of FCC and BCC structures. The NiCoFeCrGa system \cite{Huang.2017, Vida.2016, Vida.2018, Molnar.2019, Vida.2021} can be tailored on a wide scale from the cast FCC/BCC state either in terms of phase ratios or the appearance of new phases. Considering the phase transitions of the material and the fact that the ferro-paramagnetic transition takes place in a range close to room temperature in this HEA, the alloy is a good candidate for functional applications, either as a micromechanical component or as a magneto-caloric micromaterial. These applications require the understanding of mechanical properties at the micron scale. In addition, we consider this HEA as a model material for dual-phase FCC/BCC alloys, and to understand its bulk plasticity, one should investigate the mechanical properties of the distinct phases. 
Since the characteristic scale of the phase structure is $\sim 10 \, \upmu$m, experimental approaches call for tests performed at small scales with high lateral and temporal resolution. \emph{In situ} high strain rate (HSR) testing has only been established very recently at the relevant scale \cite{Guillonneau.2018}. Some industrial examples of HSR experimentation include the \emph{(i)} impact resistance of turbine blades, \emph{(ii)} studying the survival of components/devices when dropped, \emph{(iii)} crash testing, \emph{(iv)} metalworking operations, \emph{etc}. Such extreme applications require clear understanding of the rate-dependent deformation kinetics. Furthermore, providing repeatable experimental data in the HSR regime can improve existing numerical models, that cannot simulate materials properties close to the quasi-static deformation regime (i.e. molecular dynamics \cite{Fan.2021}). Additionally, by looking at a combined FCC/BCC system where the compositions of the constituent phases are as close as possible, the role of the lattice structure played in the deformation dynamics can be studied.

The local mechanical behaviour of single phase HEAs have been tested rigorously at quasi-static strain rates (QSSR,  0.0001/s -- 0.1/s)  \cite{Zou.2018} at room temperature by micropillar compression \cite{Heczel.2018}, tensile testing \cite{Alfreider.2021} and nanoparticle compression \cite{Yan.2022}. High strain rates (HSR, $>$10/s) have only been accessible so far at the macro-scale (mm), and by various simulation methods. The effect of irradiation \cite{Peng.2022}, size effects related to lamellar orientation \cite{Chen.2021}, pure FCC \cite{Raghavan.2017, Basu.2018}, BCC \cite{Zou.2014} and HCP \cite{Soler.2018} single crystal HEAs were also experimentally investigated by micropillar compression and nanoindentation \cite{Zhao.2020} at QSSR. 

The comparison between single and multiple phase materials is quite difficult, not to mention the various heat treatments, alloy compositions and microstructures. Regarding the size effects, Zou \emph{et al.} \cite{Zou.2015} reported increased yield strength at 5\% deformation with decreasing pillar sizes in refractory HEA thin films. In this case, the thin film layer has a strong texture and a columnar microstructure due to the deposition process. The hardening rate was also size dependent, as smaller pillars exhibited increased hardening at higher strains. The size effect was investigated in case of pillar diameters below 1 $\upmu$m, where FIB machining can play an important role, leaving a significant ion-modified layer on the surface of the specimen. However, the HEA in the current study is different from nanocrystalline and other pure single crystalline alloys. 

Bulk HEA was also used to fabricate pillars with different orientations \cite{Zou.2014}, and it was found that the yield stresses of micropillars are significantly larger than in case of bulk specimens. It is important to note that the studied alloy in Ref. \cite{Zou.2014} was homogenized for 7 days at a high temperature, therefore it is considered as a stabilized BCC alloy with relatively large ($>100$ $\upmu$m) grains. It was also stated that this size related effect is much smaller than in case of pure metals. Size effect is usually considered as a consequence of the stochastic presence of dislocation sources in pure metals. However, yielding in HEAs is dominated by single dislocation dynamics because of the high lattice friction. Models/simulations predicting the yield stress usually only consider the jerky motion of a single dislocation\cite{Varvenne.2016, Geslin.2021}. Therefore, in this case, collective dislocation effects do not seem to play an important role. In traditional metals, the interaction between dislocations are significant, and that determines the yield stress (e.g., Taylor hardening).

Mechanical properties not only depend on the chemical composition of the alloy but also the local non-random distributions of the constituting elements \cite{Ding.2019, Zhang.2020}. Local chemical heterogeneity can be triggered by, i.e., the speed of the cooling of the samples after alloying \cite{Vida.2021}. Individually probing each phase in these complex materials require microscale experiments to be performed, which is often done by microindentation mapping on the bulk sample's surface \cite{Muthupandi.2017, Sinha.2019}.

Experiments have already proved the efficiency of micron sized sample deformation (micropillar compression, tensile bar testing, microcantilever bending etc.), when a small fraction of the bulk volume can be shaped by focused ion beam (FIB) milling. The sensitivity to strain rate in case of various HEAs has been studied at mainly by indentation \cite{Jiao.2016} and bulk mechanical testing \cite{Kumar.2015, He.2018, Gangireddi.2018a, Gangireddi.2018b}, occasionally by micropillar compression \cite{Sadeghilaridjani.2019}. However, to our knowledge, strain rate sensitivity (SRS) has not yet been investigated at the HSR regime on a dual-phase HEA, therefore the SRS-behaviour of such complex materials is unknown.

The aim of the current study is to understand how a dual-phase system with two distinctive deformation behaviours may affect the global plastic response by characterizing SRS in the constituent phases. Dislocation interactions with phase and precipitation boundaries were also examined by state-of-the-art analytical methods. In this study we investigate size effects, strain rate effects and the phase related mechanical behaviour of a dual phase HEA by micropillar compression. More than a hundred micropillars with three different diameters were prepared and tested during the campaign. Furthermore, it is important to understand how micromechanical results can be influenced by the mixture of phases within a single micropillar under extreme conditions, hence, we aim to widen the applicability of micromechanical testing by performing compression tests in the ultra-high (up to $\sim 10^4$/s) strain rate regime.

\section{Experimental}

\begin{figure*}[!hb]
    \centering
    \includegraphics[width=0.8\textwidth]{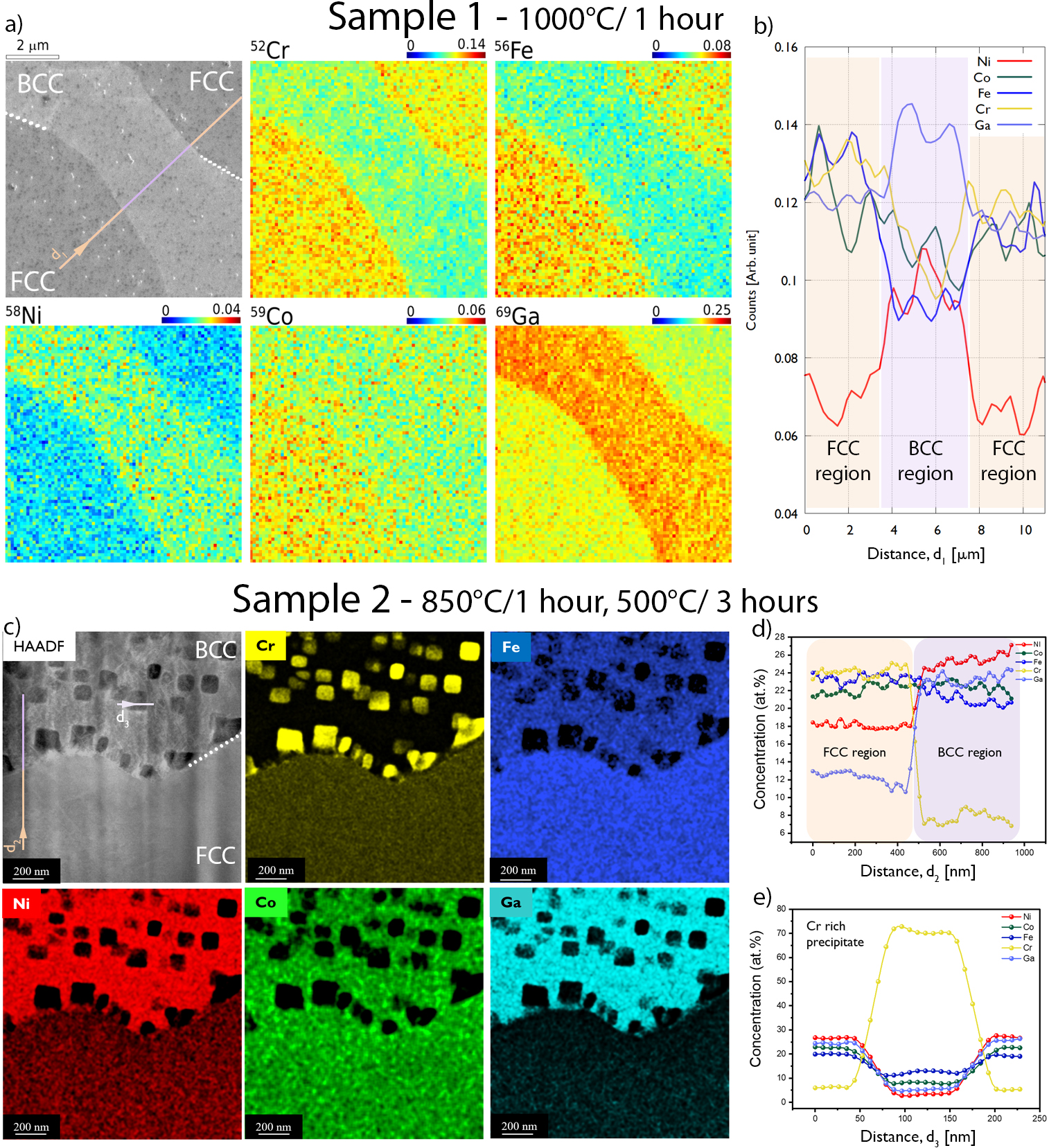}
    \caption{\textbf{Chemical composition of the samples.}  Phase boundaries are indicated by white dotted lines. \textbf{(a)} Sample 1 by TOF-SIMS (color scales are in arbitrary units: counts/TOF extraction). \textbf{(b)} Chemical profile ($d_1$) extracted from the line marked on the secondary electron inset. \textbf{(c)} Sample 2 by TEM-EDS. Chemical profiles over the \textbf{(d)} FCC/BCC ($d_2$), and \textbf{(e)} BCC/precipitate regions ($d_3$) extracted from the line marked on the high-angle annular dark field (HAADF) panel.
    \label{fig:01}}
\end{figure*}

\subsection{Sample preparation}\label{sec:sample}

\textbf{Casting, microstructure characterization.} Sample preparation was identical as reported by Vida \emph{et al.} \cite{Vida.2021}, where the basic microstructure characterization methods were performed on these materials. For the current study, two different heat treatments were employed. \emph{Sample 1 -- S1} was heated to $T_{S1}=1000^{\circ}$C for 1 hour and quickly cooled by water, while \emph{Sample 2 -- S2} was slowly cooled by the furnace ($T_{S2,I}=850^{\circ}$C for 1 hour then at $T_{S2,II}=500^{\circ}$C for 3 hours). Due to the slower cooling rate, S2 exhibited precipitates (with average diameters of $\bar{d}_\mathrm{prec}\sim 100-150$ nm) within the BCC phase, while the composition of the FCC phase remained homogeneous. In the S1 specimen, the rapid cooling did not allow the system to form nanoprecipitates in either of the constituting phases. The microstructure remained complex (dendritic FCC surrounded by BCC matrix) due to the absence of the homogenization heat treatment after casting, that forced the experimental characterization to approach the sub--5 $\upmu$m scale, making micromechanics indispensable to separate the influence of phases on the global mechanical response to deformation. Orientation and phase mapping are provided in the Suppl.\ Material.

It is crucial to understand the subtle chemical changes between different samples and composing elements, as phase boundaries and precipitates can have major effects on the mechanical response to external load due to the modified dislocation behaviour. In Figure \ref{fig:01}a, chemical analysis performed by time-of-flight secondary ion mass spectroscopy (TOF-SIMS) is shown. FIB-based TOF-SIMS measurements were conducted using a TOFWERK Ag.\ (Switzerland) detector at 30 kV 40 pA, with additional fluorine gas injected during the  sputtering process to enhance secondary ion signals. In case of S1, $^{59}$Co maps show no substantial difference between FCC and BCC regions, while in cases of $^{52}$Cr, $^{56}$Fe, $^{58}$Ni and $^{69}$Ga, TOF-SIMS detected a variance in concentration between the two phases. S1 has a more homogeneously distributed composition, where precipitates cannot form due to faster cooling following casting. FCC regions seem to have somewhat higher $^{52}$Cr and $^{56}$Fe content, while depleted of $^{58}$Ni and $^{59}$Ga ions.

In order to investigate the chemical composition of S2 in more detail, transmission electron microscopy (TEM)-based energy dispersive spectroscopy (EDS) was applied (Figure \ref{fig:01}c). It can be seen that the small cubical precipitates in S2 are rich in Cr. As they form only in the BCC phase, the matrix itself has lower $^{52}$Cr content than the FCC grains \cite{Vida.2021}. It can also be observed that precipitates often prefer to accumulate along grain/phase boundaries (see also in Suppl.\ Fig.~\ref{fig:s14}, measured by TOF-SIMS in S2 along a BCC grain boundary). The FCC phase is somewhat richer in Fe, while the BCC matrix has higher concentration of Ni and Ga, which is in good agreement with the TOF-SIMS results. In typical metal-alloy compositions, Ni and Cr play a role of $\gamma$- and $\alpha$-phase stabilizing elements, however, in this HEA, the addition of Ga is the reason why the system becomes duplex \cite{Vida.2016}, hence the compositional roles are different for these two elements. Concentration profiles along certain significant lines are plotted in Fig.~\ref{fig:01}b,d,e for comparison. 

Before FIB preparation, the surface of the specimens was grounded and polished with diamond suspension (final step of 1 $\upmu$m particle size) and with colloidal silica (Struers OP-S and OP-U). The average grain sizes of S1 and S2 are listed in Suppl.\ Table~\ref{table:s0}, determined by large area electron backscatter diffraction (EBSD) mapping. It is important to note that due to the expanding nature of dendritic grain growth, the statistical analysis of grain sizes (i.e., equivalent circle diameter) might be influenced by the mapping area.

\textbf{FIB milling.} For the micromechanical tests, circular pillars with various diameters ($\sim 1.5 - 3.5$ $\upmu$m, for more detailed information, see Supplementary Material Section \ref{sec:sspillars}) were fabricated by FIB. These sizes made it possible to fabricate numerous pillars from one grain, and also they were chosen such that studying size related effects in both constituting phases was possible.
Pillar fabrication was carried out using Tescan (Lyra3) and FEI (Helios 600i) dual beam scanning electon microscopes (SEM)  through a sequence of standard beam settings (30 kV, $10-0.04$ nA). Images taken during the various preparation steps of a 3 $\upmu$m diameter pillar are shown in Suppl.\ Fig.~\ref{fig:s1}.

During pillar preparation, the observation was made that BCC phase sputters much faster (by a factor of 2) than FCC. This effect is shown in Suppl.\ Fig.~\ref{fig:s9}. Due to this, micropillar bases might not necessary align with the surface of the sample, making it difficult to estimate the exact height of each pillar. As it is shown in the \emph{Results} section (Sec.~\ref{sec:results}), sometimes micropillars had a phase boundary within the deformed volume. This was inevitable, as phase mapping could only be performed on the surface of the bulk sample. In order to ensure that the prepared pillars would contain mostly one phase, numerous micropillars were prepared. Due to the significant differences between the FCC/BCC mechanical behaviour, distinguishing pure and mixed modes is relatively straightforward after analysing the mechanical data. Furthermore, several pillars were lifted out and their cross-sections were analyzed by TEM and high (angular) resolution EBSD (HR-EBSD) mapping. This also presented a unique opportunity to investigate dislocation interactions along the phase boundaries and in the presence of precipitations.

\subsection{Micromechanical testing}
Micropillar compression tests provide close to ideal uniaxial conditions during deformation, leading to uniform stress/strain fields, hence this technique is favoured here as opposed to nanoindentation. Pillar compression tests were carried out using a SEM-com\-pa\-tible microdeformation stage (ASA, Alemnis AG). Imaging during testing was performed with a Zeiss Supra 55VP and a Zeiss DSM 962 SEM. For quasi-static studies (SR $\leq$ 0.1/s) strain gage based load cells (standard and mini load cell -- SLC and MLC) were used to record the force acting on the micropillars and during  indentation mapping \cite{Raj.2021}. HSR compression tests were performed using a piezo-based platform (SmarTip \cite{Raj.2019, Raj.2021} and Enhanced SmarTip with extended actuation distance). High fidelity data were captured using the supported hardware and software with sampling rates of 40 Hz -- 1 MHz. An automatic approach with 0.15 mN threshold was applied to precisely detect the contact of the tip with each pillar. After the automatic surface detection, linear loading profiles were applied on the micropillars with varying strain rates (SR).  Occasionally, when the maximum programmed displacement was reached, the tip was moved backwards by 10 nm and the displacement was kept constant for 20 s (used for the thermal drift corrections). In some cases, strain rate jump tests (SRJT) \cite{Raj.2022} were performed in order to investigate SR sensitivity using the same pillar. In these tests, an initial loading with 0.001/s was followed by 0.1/s, 0.001/s, 0.01/s and 0.001/s loading segments (see in Suppl.\ Fig.~\ref{fig:s10}). Auto-approach procedures were not performed directly on the pillars tested at HSR due to the increase in load noise ($\sim 0.3$ mN). In order to reduce unwanted resonance of the diamond flat punch due to high levels of acceleration/deceleration, a ``proportional" displacement profile was also applied in some cases of HSR experiments (Suppl.\ Fig.~\ref{fig:s11}, and Suppl.\ Table~\ref{table:tb} marked with ``prop"). In every other instances, a linear loading profile with a given tip velocity was applied (Suppl.\ Tables~\ref{table:ta}, \ref{table:tb} and \ref{table:tc}).


Load-displacement curves of micropillars and nanoindentations were compliance corrected and analyzed by the Alemnis AMMDA v2.01 software. For the Young's modulus and hardness calculations, the unloading part of the load-displacement curves were used, along with $\overline{\nu}=0.31$ Poisson's ratio. More information about the calculation of $\overline{\nu}$ can be found in Suppl.\ Section \ref{sec:ss2}.  FCC and mixed phase pillars usually showed a smooth elastic loading part followed by a jerky plastic part having significant stress drops characteristic to the presence of the FCC phase. In these examples, the maximum stress at the end of the elastic part (before the first main stress drop) was defined as the yield stress ($\sigma_\mathrm y$). During high strain rate (HSR) testing, due to the high acceleration of the tip (reaching up to some tens of mm/s actuation velocity), an oscillating noise appears in the signal (``ringing effect"). In order to determine $\sigma_\mathrm y$ in such cases, the elastic and plastic parts of the stress-strain curves were fitted using a linear function, and their intersection was used to obtain the apparent yield stress values \cite{kalacska.2024}. To make the evaluation consistent at all applied strain rates, $\sigma_\mathrm y$ values were obtained using the same yield criterion in case of low strain rate (LSR) pillar testing too, where the transition between the elastic and plastic part of the curve was smooth.

\subsection{Analytical Methods}

\textbf{EBSD maps} (Oxford Instruments Nordlys Nano and Symmetry 2) were recorded prior to and after pillar fabrication to choose grains with similar orientation. All pillars were oriented to single slip (pillar orientation and phase maps can be found in Suppl.\ Figs.~\ref{fig:s2}--\ref{fig:s4}), depending on the available grains. Grain statistics were collected using AZtecCrystal v2.2.

\textbf{HR-ESBD} evaluation was performed using BLGVantage CrossCourt Rapide v4.6 software. The cross-sections of lifted out micropillars were prepared using a 30 kV 0.6 nA Ga$^+$ beam, then diffraction patterns were recorded with a 20 kV, 7 nA electron beam and an image frame averaging of 5. Some pillars were prepared for TEM analysis (see Suppl.\ Tables~\ref{table:ta}, \ref{table:tb} and \ref{table:tc}) that were later studied by high resolution transmission Kikuchi diffraction (HR-TKD) using a 30 kV electron beam for geometrically necessary dislocation (GND) density imaging. During the evaluation, the following elastic constant coefficients were used [determined for a ferromagnetic material, in GPa] for BCC: $C_{11}^\mathrm{BCC} = 215.1$, $C_{12}^\mathrm{BCC} = 162.7$, $C_{44}^\mathrm{BCC} = 122.3$, and for FCC: $C_{11}^\mathrm{FCC} = 184.7$, $C_{12}^\mathrm{FCC} = 146.9$, $C_{44}^\mathrm{FCC} = 123.3$ \cite{shuo.2017}.

\textbf{Chemical analysis} of the two samples were carried out using TOF-SIMS (TOFWERK AG). This technique uses the FIB beam to sputter away locally the sample \cite{Dunn.1999, Giannuzzi.2004}, while the detector collects and categorises the sputtered ions based on their mass. The method was chosen instead of conventional SEM-based energy dispersive spectroscopy (EDS) due to its excellent spatial (lateral and depth) resolution, as sputtered ions originate only from a top few atomic layers on the surface. It is important to note that chemical composition and crystallographic orientation influences the sputtering speed, and in case of a multiphase alloy, it would mean that the sputtered crater depth varies in the FCC/BCC regions \cite{Pillatsch.2020}. Maps were recorded in ``+" mode, using a 30 kV, 40-160 pA ion beam with fluorine inert gas injected through a gas injection system (GIS, Orsay Physics). Fluorine was applied to increase ionization probability and enhance secondary ion signals during the analysis \cite{Priebe.2019}. 10-20 frames were recorded over areas of 5-10 $\upmu$m, with a dwell time of 32 $\upmu$s and $2 \times 2$ binning.

\textbf{STEM}, high-angle annular dark field (HAADF) images and selected area electron diffraction (SAED) patterns were acquired using a Themis 200 G3 aberration (probe)-corrected TEM (ThermoFischer) operating at 200 kV, equipped with a Nanomegas EDS system. Fast Fourier Transformed (FFT) images are generated from lattice fringes.

\section{Results}
\label{sec:results}

The differences between the two samples (S1 and S2) were already introduced in Section \ref{sec:sample}, and shown in Fig.~\ref{fig:01}. Based on the chemical profiles measured by TOF-SIMS and TEM/EDS, it can be concluded that the composition of the FCC phase is not influenced by the cooling rate, while S1 differs from S2 in terms of the Cr-rich precipitates in the BCC matrix. Therefore, the most pronounced dissimilarities within the global mechanical response in these two materials are directly linked to the precipitates within the BCC phase. This effect can be further studied in case of the emerging size effects. As to the variance between the FCC/BCC small scale local mechanical properties, sample S1 was chosen to be the material where size and strain rate related effects are both discussed in this work, as the evaluation of such a rapidly cooled system is missing from the literature.

\begin{figure*}[!ht]
    \centering
    \includegraphics[width=0.99\textwidth]{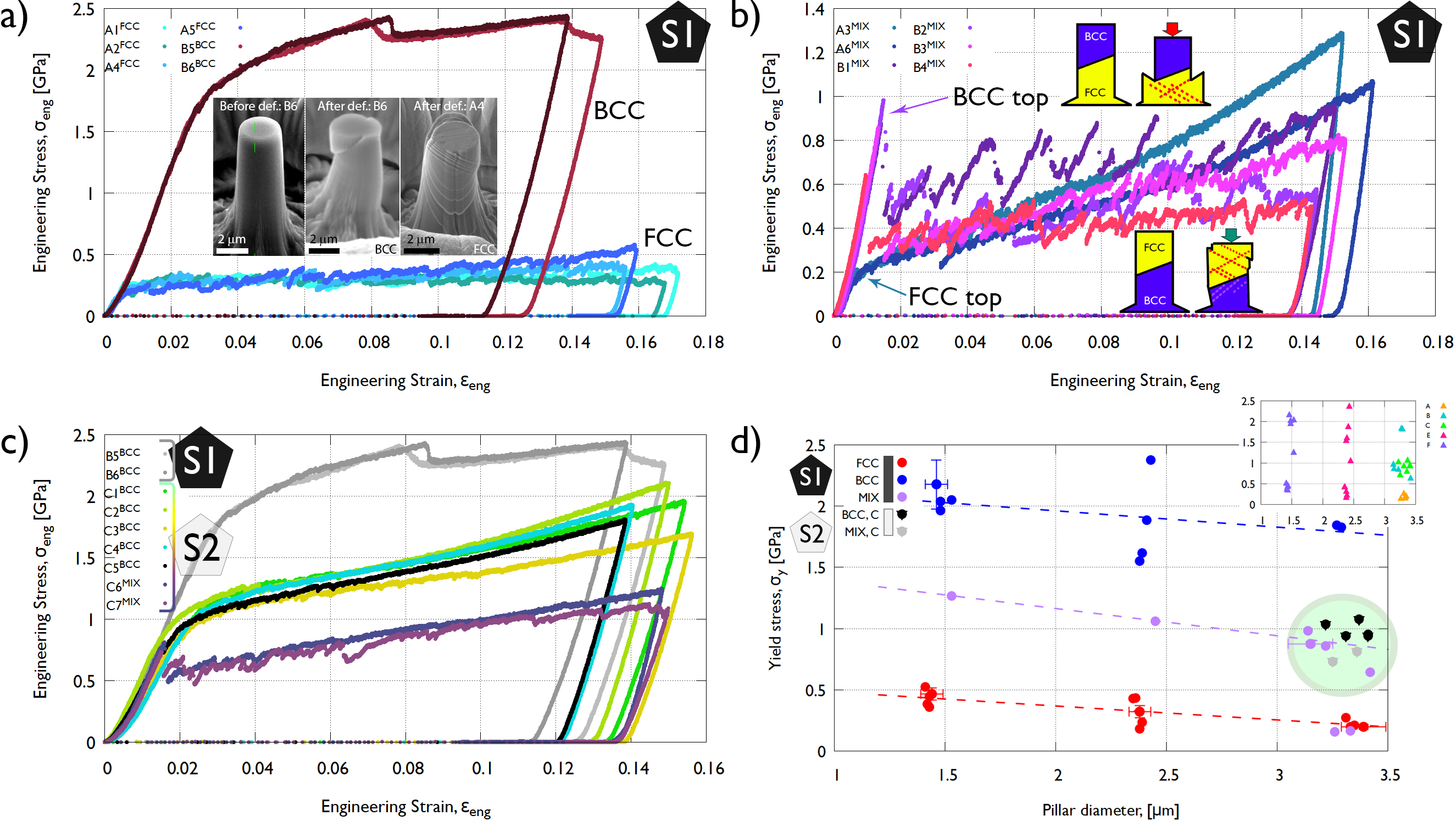}
    \caption{\textbf{Stress-strain plots at quasi-static SR} (d$\sim 3$ $\upmu$m). \textbf{(a)} Pure FCC and BCC $\sigma_{\mathrm{eng}} - \varepsilon_{\mathrm{eng}}$ comparison, Sample 1. Inset images show the pillars before and after deformation. \textbf{(b)} Pillars with mixed phases, Sample 1. Insets showing the deformation mechanisms based on imaging mixed mode pillars after compression (see in Suppl.\ Fig.~\ref{fig:s12} and \ref{fig:s13}).  \textbf{(c)} Deformation curves measured on BCC pillars on Sample 1 (S1: B5, B6) and Sample 2 (S2: C1-C7). \textbf{(d)} Yield stress dependency of the pillar diameter (\emph{``size effect"}). The inset shows which point belongs to which set of pillars in Suppl.\ Tables~\ref{table:ta} and \ref{table:tb}. 
    \label{fig:02}}
\end{figure*}

First, micropillars of $\sim$3 $\upmu$m diameter were compressed and the analysed data showed a huge variety of stress-strain curves. The results can be categorized into two main groups, namely deformation curves of \textit{(i)} pure BCC or FCC pillars, and \textit{(ii)} mixed pillars, containing a phase boundary within the deformed volume. In Fig.~\ref{fig:02}a, the $\sigma_{\mathrm{eng}} - \varepsilon_{\mathrm{eng}}$ curves show a very distinctive response to external loading.

In the BCC phase (pillars B5 and B6), the yield stress reaches extremely high values after the elastic loading ($\varepsilon \sim 0.03$), that is followed by a rather smooth plastic regime. Around $\varepsilon \sim 0.08$, a load drop of $\sim 0.2$ GPa indicates the activation of a slip plane, where the deformation continues, reducing the hardening rate afterwards. In pillar B5, a second slip is activated close to the maximum applied strain ($\varepsilon \sim 0.14$), while B6 is unloaded before this event. 

Contrary to the BCC phase, the FCC pillars show much lower yield point ($\sim$$200-300$ MPa), that is followed by a jerky plastic part. The load drops indicate dislocation avalanches activated on the possible slip planes, that are confirmed by the \textit{post mortem} imaging of the pillar surfaces meshed with numerous steps on their facets. Multiple slip planes are activated as a result of the achieved high maximum strains, that continuously produced dislocations to accommodate the shape change of the pillar. Secondary electron images of pillars before and after deformation are shown in the inset of Fig.~\ref{fig:02}a. The recorded pure deformation curves exhibit good repeatability, however, due to the small grain size (rapid cooling of S1), the microstructure holds a challenge to be evaluated by such large pillars, as phase boundaries can be included in the FIB-milled volumes. The presence of such boundaries are hidden until the pillars are deformed, or their cross-sections are analysed by another microstructure sensitive technique. As the latter was not possible in the current study before performing the compression tests, specimens were fabricated in larger numbers.

Pillars falling into the ``mixed" category are plotted in Fig.~\ref{fig:02}b. Here, two characteristic behaviour can be observed. On the one hand, samples with BCC top / FCC bottom exhibit about 3-4 times higher yield point than pillars with reverse composition and the plastic regime is dominated by load drops characteristic to the FCC phase. These load drops are due to strain burst being activated in the lower pillar volume. On the other hand, the strengths of the FCC top / BCC bottom pillars reach much higher values than previously seen, where stress drops are mainly apparent at the onset of plasticity and are relatively small. These results lead to the sketches shown in the insets of Fig.~\ref{fig:02}b, describing how mixed pillars deform under compression. Mixed pillars with BCC tops show no (or only very limited) slip activity in the upper region, while the bottom FCC part becomes dense in slip bands due to the dislocation activity (see Suppl.\ Fig.~\ref{fig:s12} red arrow and \ref{fig:s13}a. Pillars with FCC tops on the other hand start to slip immediately after elastic loading on the favoured slip planes. Generated dislocations begin to pile up against the phase boundaries, inducing a large strain field, that eventually can create new dislocations within the BCC lower region if the deformation is high enough (see Suppl.\ Figs.~\ref{fig:s12} green arrow and \ref{fig:s13}b.

\begin{figure}[!ht]
    \centering
    \includegraphics[width=0.49\textwidth]{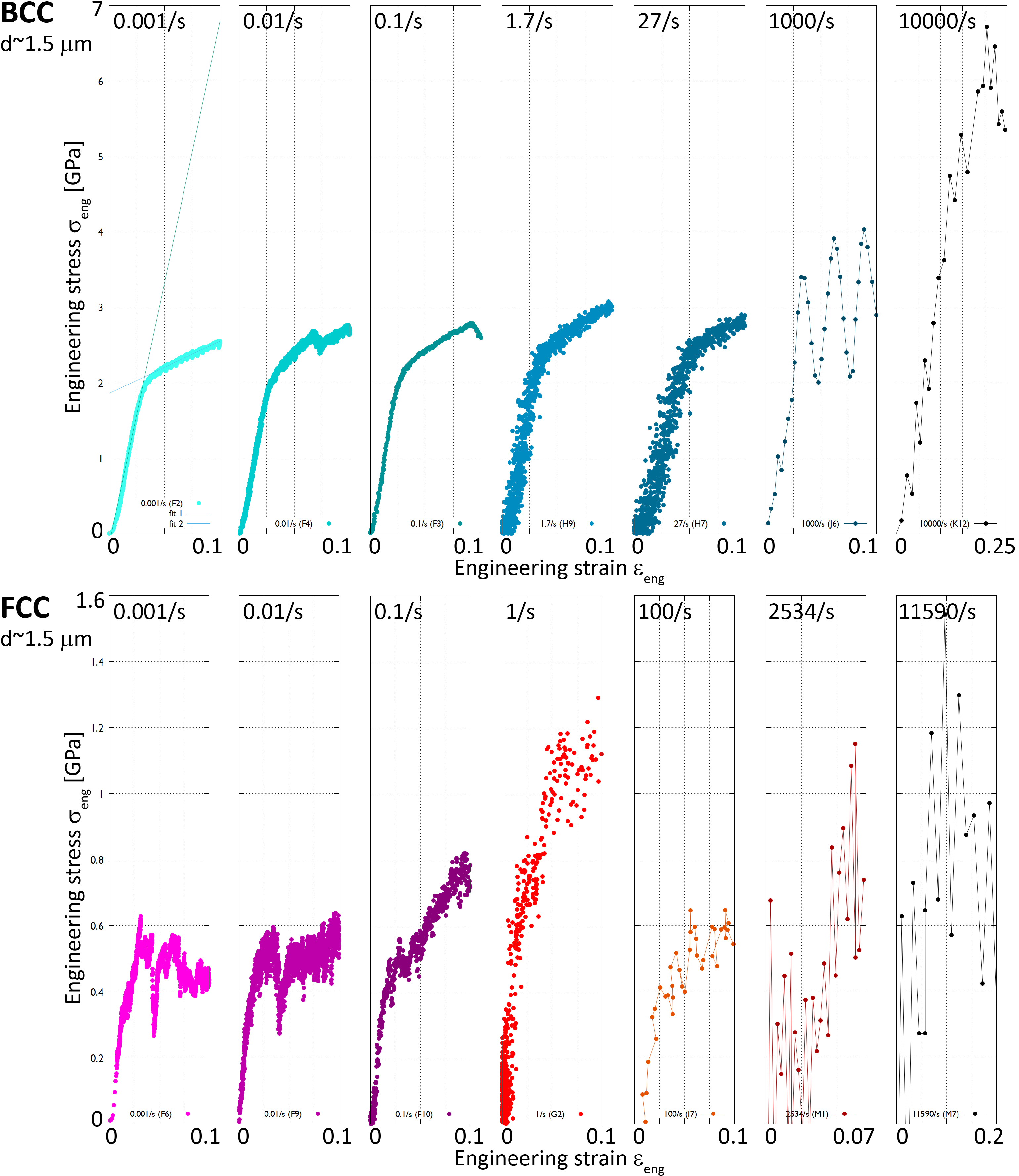}
    \caption{\textbf{Stress-strain curves} at various strain rates, Sample 1.}  
    \label{fig:03}
\end{figure}

The comparison of $\sigma_{\mathrm{eng}} - \varepsilon_{\mathrm{eng}}$ curves belonging to the S1 (BCC) and S2 (BCC and mixed: BCC top) samples are shown in Fig.~\ref{fig:02}c. In S2, the precipitates appearing in the BCC phase cause the high entropy mixture to reduce its yielding by $\sim$$50$\%. By plotting the yield stress values as a function of the pillar diameters in Fig.~\ref{fig:02}d, one can observe that the S2 sample yields at similar values as S1 mixed pillars (green highlighted area) in case of the largest tested pillars. If one looks at how the pillar size affects the yielding, a monotonous trend can be concluded, where the smaller pillars exhibit somewhat higher $\sigma_{\mathrm{y}}$. As these samples are only minimally influenced by their size, pillars with the smallest diameters (but still above the regime where FIB-induced damage would be considerable) were chosen to investigate the strain rate sensitivity behaviour. This reduces the possibility to create and analyse mixed pillars. In Fig.~\ref{fig:02}d, the $\sigma_\mathrm y$ values for the mixed pillars were set as the maximum yield stress before the first load drop. In case of the other pillars, a linear fitting method (on both the elastic and plastic parts) were considered (see in Ref.~\cite{kalacska.2024} and in Fig.~\ref{fig:03} on the BCC $\dot{\varepsilon}=0.001$/s curve) to be consistent with the more challenging high strain rate evaluation.

In Fig.~\ref{fig:03}, a few representative stress-strain curves of compressed micropillars are plotted, measured only on S1. It can be observed that with increasing strain rate, the noise on the experimental plots increase. This can be attributed to the application of various load sensors designed to operate under certain conditions. The oscillations appearing $\dot{\varepsilon} \sim 1000$/s is related to the ringing effect.

\begin{figure}[!ht]
    \centering
    \includegraphics[width=0.49\textwidth]{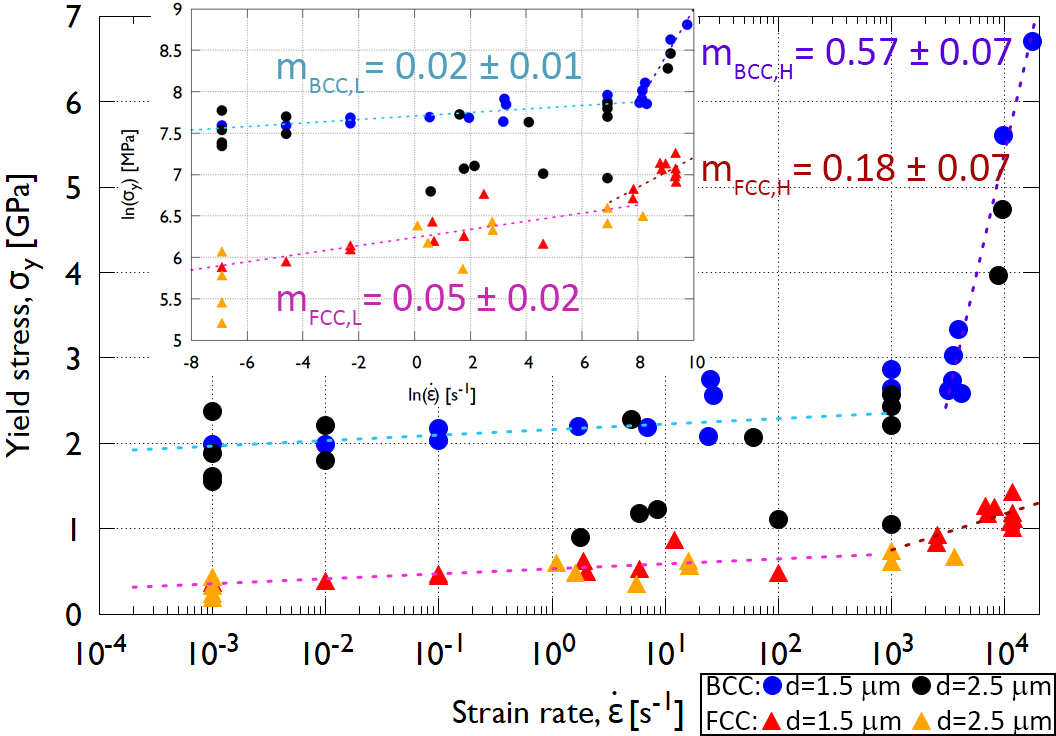}
    \caption{\textbf{Strain rate sensitivity:} The yield stress plotted as a function of the applied strain rate during compression, Sample 1, linear scaling. The inset shows the same dataset in log-log scale, that was used for the determination of the various $m$ values.
    \label{fig:04}}
\end{figure}

By extracting the yielding events from the $\sigma_{\mathrm{eng}} - \varepsilon_{\mathrm{eng}}$ curves, the SRS behaviour can be created in Figure \ref{fig:04}, plotted in log-log scale for easier comprehension. The inset shows the log-log plot of the same dataset. Following the well-established strain rate sensitivity ($m$, \cite{Caillard.2003}) calculation based on Eq. \ref{eq:srs},

\begin{equation}\label{eq:srs}
    m=\frac{\partial (\ln{\sigma})}{\partial (\ln{\dot{\varepsilon}})},
\end{equation}
the SRS factors at each relevant interval (L: low strain rates, H: high strain rates) can be extracted for both BCC and FCC phases. For the construction of the SRS sensitivity behaviour, pillars with both 2.5 $\upmu$m and 1.5 $\upmu$m diameters were used.

\begin{figure*}[!ht]
    \centering
    \includegraphics[width=0.95\textwidth]{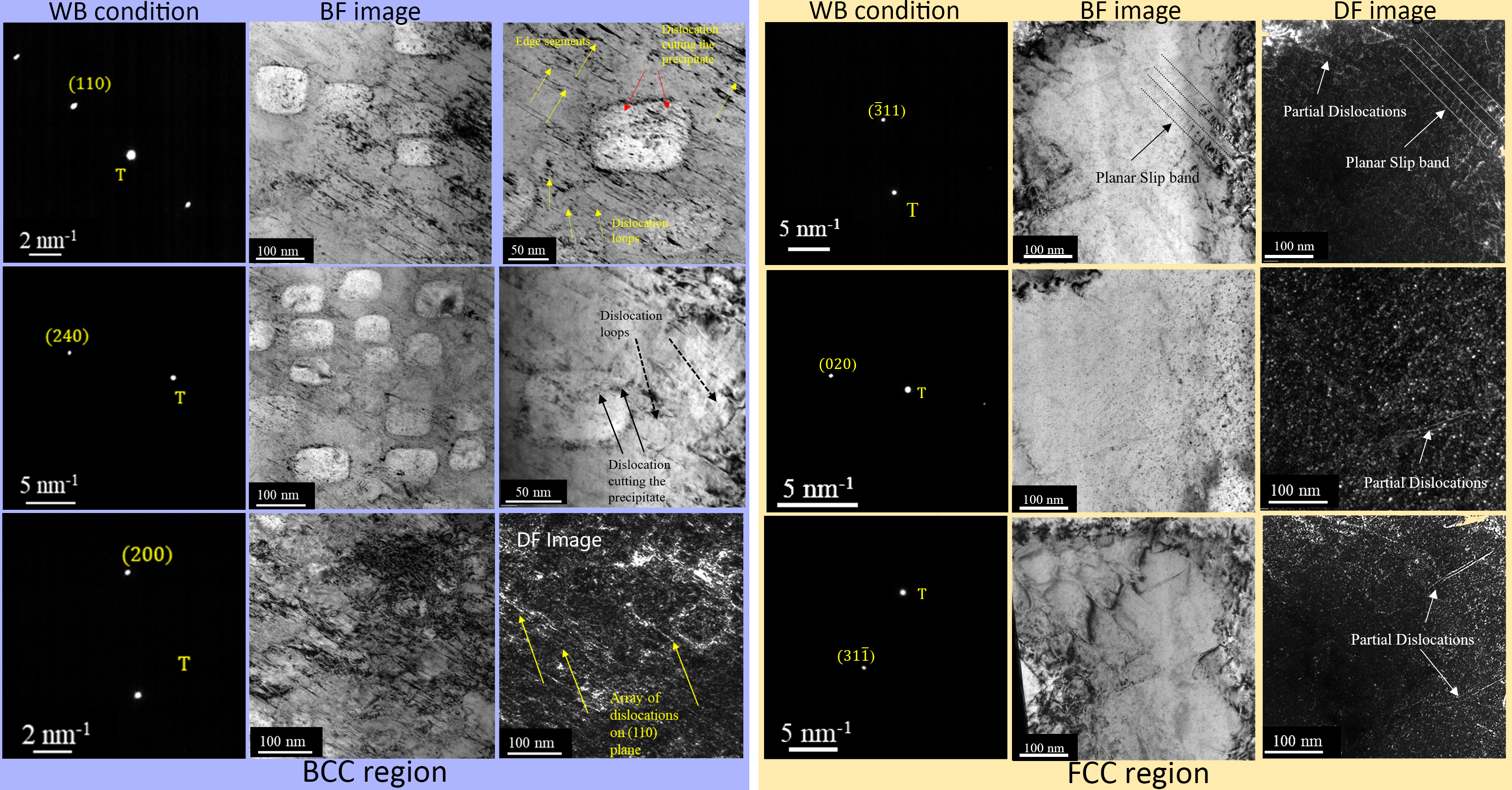}
    \caption{\textbf{TEM, dislocations} after deformation of pillar C1 (Sample 2).  
    \label{fig:05}}
\end{figure*}

\begin{figure}[!ht]
    \centering
    \includegraphics[width=0.49\textwidth]{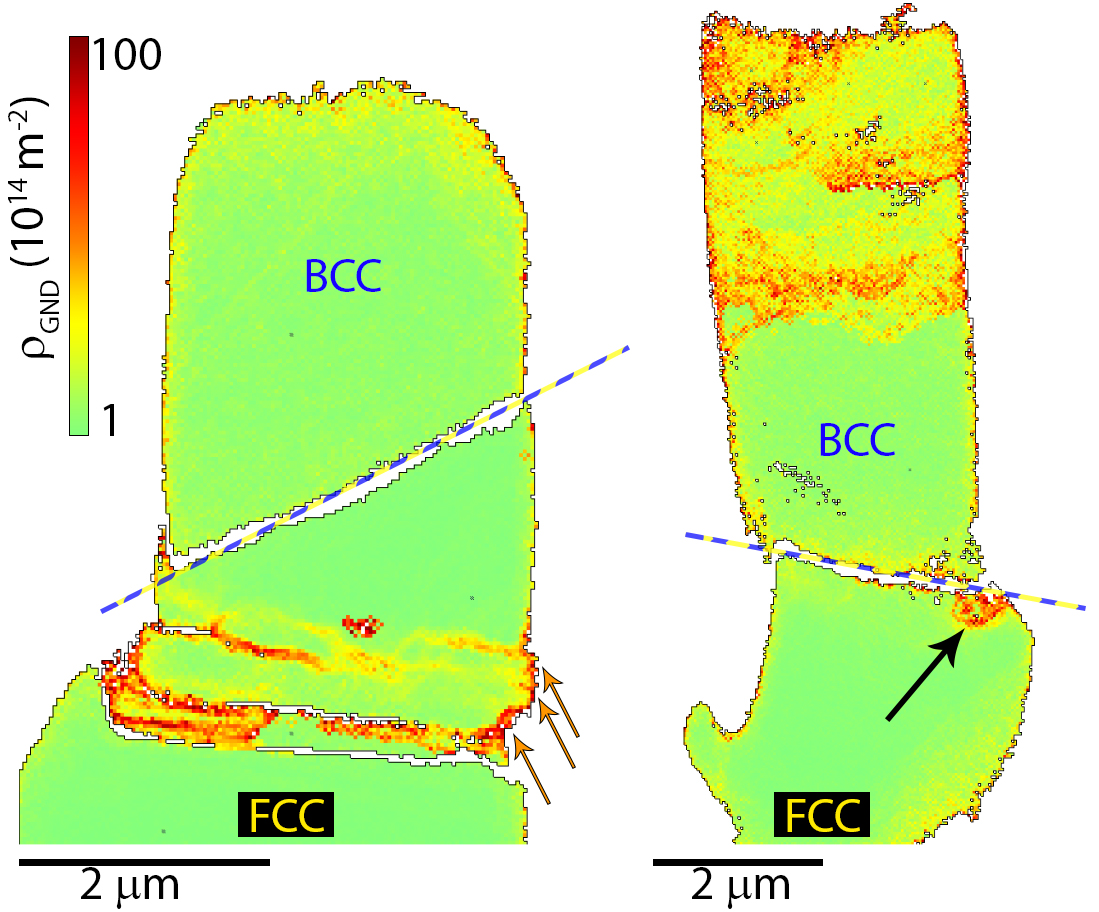}
    \caption{\textbf{HR-EBSD results,} Sample 1, cross-sectional GND density distributions measured on pillars (left) B3 and (right) B6. Conventional EBSD results are additionally shown in Suppl.\ Fig.~\ref{fig:s10} and \ref{fig:s11}. 
    \label{fig:06}}
\end{figure}

After deformation, several pillars were lifted out for cross-sectional TEM analysis. In order to understand how dislocations interact with the precipitates forming in S2, dislocation characterization by bright field (BF) and dark field (DF) imaging were concluded in Figure \ref{fig:05} and in Suppl.\ Fig.~\ref{fig:s15}. The TEM lamella was then mapped with HR-EBSD (transmission Kikuchi diffraction, HR-TKD) in the BCC phase, where Cr-rich cuboids are present. The consequent elastic strain ($\varepsilon_{ij}$) and rotation tensors ($\omega_{ij}$) are plotted in Suppl.\ Fig.~\ref{fig:s20}, while a Burgers vector analysis (\cite{kalacska.2020, kalacska2020_3d}) was performed in Suppl.\ Fig.~\ref{fig:s21}. Pillars fabricated from S1 and deformed at various strain rates are also analyzed in Suppl.\ Fig.~\ref{fig:s16}. Some pillars extracted from S1 were measured by HR-EBSD to confirm the constituent phases and to perform GND density and individual stress tensor mapping. These results are shown in Figure \ref{fig:06}, Suppl.\ Fig.~\ref{fig:s7} and \ref{fig:s8}. The surface of Sample 2 was also mapped by HR-EBSD before deformation in order to determine preexisting stress localization and dislocation density (shown in Suppl.\ Fig.~\ref{fig:s17} and \ref{fig:s18}). A typical diffraction pattern in Suppl.\ Fig.~\ref{fig:s19} highlights the good surface quality prior to pillar fabrication in Sample 2.

\section{Discussion}

In this work, the mechanical effects of the FCC and BCC phases (including precipitations) were studied at the micron scale, as the HEA's response to strain rates and sample sizes vary considerably. Two different heat treatments were used to create two distinct microstructures of the same compositionally complex alloy (S1 and S2).

Firstly, the mechanical response of two constituent pha\-ses were analyzed in S1 (FCC and BCC, Figure \ref{fig:02}). Although the two phases were created from the same molten alloy (see in Fig. \ref{fig:01}), there is a slight difference in their chemical composition (BCC is richer in Ni and Ga, but somewhat depleted of Cr), that is responsible for the variation in the final crystal structure. The BCC phase appears to be much harder than the FCC. If we apply longer heat treatment on the molten alloy (S2), the Cr in the BCC phase precipitates, that produces a BCC matrix with almost no residual Cr ($\sim 6$\% based on TEM-EDS, hence the configurational entropy decreases in this phase), that further reduces the yielding point. From an engineering point of view, this decrease in $\sigma_\mathrm{y}$ is unfavourable in terms of performance (resistance to plastic deformation) and durability. However, understanding the mechanical characteristics of the constituents is key for phase engineering, that is especially consequential in case of mixed compositions with severely different properties. In our case, looking at the stress-strain responses, we can predict the composition of the deformed pillars without prior phase analysis.

\subsection{Size effects}

Size effects shown in Figure \ref{fig:02}d seems to have no relation to the original crystal structure in S1 at the investigated length scale: both BCC and FCC have similar monotonous yield stress increase tendency towards smaller pillar sizes. Even though pillars were fabricated from differently oriented grains ([114], [104], [103], [102]), the results does not indicate a strong yield strength variation (see the inset of Figure \ref{fig:02}d for pillar series identification). Usually the slight increase in $\sigma_{\mathrm{y}}$ with the decrease of the pillar diameter can be explained by the presence of significant dislocation density present in the deformed volume, that makes the defect source density high enough to prevent a stochastic response. This could also be associated to a low scattering of the $\sigma_{\mathrm{y}}$ values no matter the sample size. To verify this, GND density measurements by HR-EBSD were carried out on the pristine material (shown in Suppl.\ Fig.~\ref{fig:s17}). The mapping performed in S2 around phase boundaries showed similar and relatively high initial dislocation density in both FCC and BCC areas ($\overline{\rho_{\mathrm{GND}}^{\mathrm{BCC}}}=1.05 \times 10^{14}$ m$^{-2}$, $\overline{\rho_{\mathrm{GND}}^{\mathrm{FCC}}}=8.71 \times 10^{13}$ m$^{-2}$). Looking at the individual stress tensor components (Suppl.\ Fig.~\ref{fig:s18}) it can be concluded that significant stress concentration ($\sigma_{11}$, $\sigma_{22}$) is located in the vicinity of the phase boundaries. Furthermore, the Cr-precipitation in the BCC section also appears to be acting as hotspots of such stresses. Local stresses (see i.e. $\sigma_{12}$ and von Mises stress maps) can further modify dislocation behaviour at boundaries, that likely contribute to the detected $\sim$ 50\% decrease in yielding of S2 compared to the fast cooled S1 material, since less external load is required for dislocation activation due to already existing residual internal stresses. 

The fitted size effect sensitivity values (from the dotted lines in Figure \ref{fig:02}d) are: $\gamma^{\mathrm{BCC}} = -0.13 \pm 0.11$, $\gamma^{\mathrm{FCC}} = -0.11 \pm 0.02$, $\gamma^{\mathrm{MIXED}} = -0.13 \pm 0.11$, that are considerably smaller than in other reported HEAs ($\sim 0.2-0.3$) \cite{Zou.2014, Xiao.2020}. This is a clear indication that homogenization through extended heat treating will also increase the sensitivity of the material to the size of the sample. Earlier reported size effects in HEA micropillar compression tests show inconsistent results. In the range of 1-10 $\upmu$m a power law relation with a log-log exponent of -0.32 was determined \cite{Raghavan.2017} in FCC single phase HEA, which is less significant than in pure FCC metals ($\sim -0.6$) at high strain rates \cite{Zou.2018}. Here, in both BCC and FCC (and mixed) pillars, the $\gamma$ parameter is the same within the error range, however, more scatter is detected in the BCC $\sigma_{\mathrm{y}}$ values.

\subsection{Strain rate sensitivity}

By studying the $\sigma_{\mathrm{eng}}-\varepsilon_{\mathrm{eng}}$ curves in Figure \ref{fig:03} and Suppl.\ Fig.~\ref{fig:s9}, one can conclude increased strain hardening rates in the FCC phase with increasing $\dot{\varepsilon}$. Furthermore, the yielding point is typically reached at higher applied strains in both cases in the HSR regime. The oscillations caused by the ringing effect makes it more challenging to extract the exact $\sigma_{\mathrm{y}}$ values, but the applied linear fitting methods produces consistent and repeatable results. SR jump testing (SRJT) has its own limitations (Suppl.\ Fig.~\ref{fig:s10}), as the applied sampling rate cannot be too different for the studied strain rates, since only one load cell is used during measurement, that is specific to a certain SR regime. SRJT performed on 2 $\upmu$m diameter pillars (series E) confirmed that these pillars (FCC or BCC) does not show strong strain rate sensitivity (Suppl.\ Fig.~\ref{fig:s10}c), that is in agreement with the conventional constant $\dot{\varepsilon}$ experimental results. This is expected from the literature \cite{Raghavan.2017}, where it is shown that HEAs typically exhibit less significant strain rate effects than pure metals, while this behaviour can be attributed to the larger lattice distortion, which affects the dislocation motion through the lattice friction (Peierls) stress.

After extracting the $\sigma_{\mathrm{y}}$ values and creating the SRS plot in Figure \ref{fig:04}, the difference between the LSR and HSR regime behavior of the two phases are evident. In the BCC phase, the SRS factor exhibits a sharp increase above $10^3$/s, transitioning from $m_{\mathrm{BCC,L}} = 0.02 \pm 0.01$, $\dot{\varepsilon} \subset [10^{-3},10^{3}]$ to a compelling $m_{\mathrm{BCC, H}} = 0.57 \pm 0.07, \dot{\varepsilon} \subset [3\times 10^{3},2 \times 10^{4}]$, that is an approx. 28 fold increase. Similar behaviour has already been reported in bulk UHSR experiments \cite{Follansbee.1988, Rittel.2006, Kumar.2015} and by simulations \cite{Fan.2021}, but not on the micro-scale before. The significance of performing such an experiment is to finally bridge experimentally (high) and computationally (low) achievable strain rates together on a similar length scale (i.e by discrete dislocations dynamics -- DDD, or molecular dynamics -- MD modelling). Looking at the FCC SRS factor, a less momentous ($\times3.6$, about 8 times weaker than in BCC) transition occurs above 10$^3$/s SR, from $m_{\mathrm{FCC,L}} = 0.05 \pm 0.02, \dot{\varepsilon} \subset [10^{-3},10^{3}]$ to $m_{\mathrm{FCC,H}} = 0.18 \pm 0.07, \dot{\varepsilon} \subset [10^3,2 \times 10^4]$. It is noteworthy that due to the generally lower yielding in this phase, the extraction of $\sigma_{\mathrm{y}}$ values is much more challenging. However, the multiple repetitions gave consistent reading on the final $\sigma_{\mathrm{y}}$ even when $\dot{\varepsilon}>10^4$. The reason behind such transition in the SRS factor at  HSR is generally considered to be the fact that in this deformation regime, dislocation movement is no longer only controlled by thermal activation (such as at LSR and low applied stresses) \cite{Salvado.2017}. At HSR, dislocation motion is mostly drag controlled, that leads to the observed overdamped defect mobility as a result of interactions with lattice phonons and electrons \cite{Lerma.2016}. Consequently, the viscous drag force will be proportional to the speed of dislocation in this regime \cite{Lerma.2014}, meaning that higher forces will be required to initiate dislocation motion upon loading, resulting in the increased yielding point.

It is worth mentioning that a few pillars that were originally classified as BCC with 2.5 $\upmu$m diameter showed much lower yielding than expected ($\dot{\varepsilon} \subset [10^{0},10^{3}]$) as a result of being compositionally mixed (BCC top, FCC bottom), but they are still plotted in Figure \ref{fig:04}. These pillars were excluded from the $m$ determination.

There is no clear difference in the SRS between the larger (2.5 $\upmu$m) and smaller (1.5 $\upmu$m) pillars, however, as the larger pillars often contained a phase boundary, it was difficult to extract values for the pure phases (even when pillars were tested in such high numbers). The size reduction must certainly play a role in terms of the number of initially present dislocations. In the current work, it was impossible to study this effect due to the complex nature of the material and the challenging micromechanical testing. On the other hand, other groups have already looked into the theoretical background of dislocation density affecting the SRS of yielding, i.e. Fan et al. \cite{Fan.2021}. Researchers have applied DDD and MD simulations to predict the yielding of pure FCC (as in Cu single crystal) as a function of initial dislocation density. They have also identified a regime around $10^3$/s, where $\sigma_{\mathrm{y}}$ increases substantially. 


Based on these simulation results, we found our experimental outcome to be consistent with the theoretical prediction, namely that higher initial dislocation density (as expected in larger pillars) results in less significant SRS increase than in samples containing more mobile dislocations. This can be observed in Figure \ref{fig:04}, where the $\sigma_{\mathrm{y}}$ values extracted from bigger pillars (black dots) are somewhat smaller than in case of the 1.5 $\upmu$m pillars (blue dots) above 10$^3$/s. The $\sigma_{\mathrm{y}}$ ratios are in accordance with the size effect measurements at lower strain rates, meaning that this phenomena is preserved (and monotonous) within the studied strain rate range.

One possible cause of the smaller SRS increase in FCC could be related to the difference in the initial dislocation density in this phase, therefore dislocation analysis by TEM was performed on compressed pillars of Sample 1 (shown in Suppl.\ Fig.~\ref{fig:s16}) and in Sample 2 (in Figure \ref{fig:05} and Suppl.\ Fig.~\ref{fig:s15}). Multiple pillars deformed to a lower maximum strain level were lifted out, that confirmed the presence of large amounts of dislocations within the investigated samples. Hence individual Burgers vector analysis was not possible by TEM in Sample 1.

The analysis of Sample 2 by TEM was more fruitful. Generally speaking, the appearance of precipitates could manifest in increased yielding in alloys (i.e. precipitation hardening). In our case, the Cr-rich cuboids in the BCC matrix induce the opposite: the yield stress values drop drastically in S2. As we look at the dislocation interactions with these precipitates in Figure \ref{fig:05}, it can be observed that dislocations simply cut through the cuboids instead of being pinned down at the interfaces. Dislocations (both straight sections and loops) were observed using weak-beam (WB) diffraction conditions (bright field -- BF, dark field -- DF) by TEM in a micropillar (C1) compressed at LSR. HR-TEM analysis at the matrix-precipitate interface confirmed the coherent nature of the boundary (Suppl.\ Fig.~\ref{fig:s15}), that facilitates dislocation crossing (hence the missing precipitation hardening effect). Contrary to an earlier TEM investigation performed on the same material \cite{Vida.2021}, the extra diffraction spots corresponding to the Cr-rich cuboids were clearly identified on the fast Fourier transformed (FFT) imaging. It is interesting to mention the detection of lattice distortions in the HR-TEM matrix image, similarly to Ref. \cite{Zou.2014}.

Some areas exhibited an array on dislocations on the (110) slip plane in the BCC region. The sample contained a phase boundary close to the bottom, that allowed TEM analysis on the FCC region too. In this phase, dislocation pile-ups were observed on planar slip bands, forming a more organized defect network. The ability of dislocations to freely move within this phase contributes to the decrease of $\sigma_{\mathrm{y}}$ and allows defects to escape the pillars though the surfaces. Partial dislocations were also detected in the FCC region, that is expected due to the generally lower stacking fault energy of FCC phase compared to BCC.

In order to have some information on the collective Burgers vectors of the generated defects withing the BCC phase, HR-EBSD mapping was performed on the TEM lamella (also referred to as HR-TKD) of pillar C1. The resulting elastic strain ($\varepsilon_{ij}$) and rotation tensor ($\omega_{ij}$) elements are plotted in Suppl.\ Figs.~\ref{fig:s20} and \ref{fig:s21}. It can be seen that the presence of precipitation (circled areas in Suppl.\ Fig.~\ref{fig:s21}a) causes local strain variation both in the normal and shear strain maps. Components of the Nye dislocation density tensor ($\alpha_{i3}$) can be calculated from the elements of the distortion tensor $\beta_{ij} = \partial_j u_i$, with $\bm u$ being the displacement field (accessible by HR-EBSD) as \cite{Wilkinson2009, kalacska.2020}:
\begin{eqnarray}\label{eq:ai3}
\alpha_{i3} = \partial_1 \beta_{i2} - \partial_2 \beta_{i1}, \quad i=1,2,3.
\end{eqnarray} 
The resulting maps of $\alpha_{13}$ and $\alpha_{23}$ correspond to edge type dislocations, while the third component ($\alpha_{33}$) shows screw type of GND features. Based on the HR-TKD analysis one can conclude that the majority of dislocations detected in the lamella were edge types, which is in agreement with the HR-TEM findings.

\subsection{Microstructure effects}

Two pillars (B3 and B6) from Sample 1 were further subjected to cross-sectional HR-EBSD analysis following the LSR compression experiments (Figure \ref{fig:06}). From the GND density maps we can deduce that in case of a mixed composition within the pillar (BCC top, FCC bottom, B3), the hard BCC remains mostly dislocation free (however, a slight increase close to the flat punch / pillar interface can be detected).  Upon reaching a critical external loading, the bottom FCC ``collapses", producing larger amount of GNDs that easily move within this section, creating slip traces visible on the surface (marked with orange arrows). Experimental evidence was already provided in the literature for the tendency of FCC having more mobile dislocations (screw dislocations in BCC are harder to activate), resulting easier annihilation upon reaching the pillar surfaces \cite{Greer.2006, Kiener.2011}. As a result, there is the difference in FCC/BCC hardening rates in the $\sigma_{\mathrm{eng}} - \varepsilon_{\mathrm{eng}}$ plots (see in Suppl.\ Fig.~\ref{fig:s9}). To the contrary, BCC materials exhibit a dislocation self-multiplication mechanism \cite{Greer.2008, Weinberger.2008}, that causes defects to multiply and form an entangled dislocation network before leaving the system. This difference is observable in the two pillars in Figure \ref{fig:06}, where GNDs form clear line structure, whereas the BCC GND network is more homogeneously distributed.

On the other hand, if the pillar was pure BCC along the whole thickness (B6, where the phase boundary is located just below the base of the pillar), the BCC phase shows increased GND activity in its upper section. However, as this part is pushed into the softer FCC base, dislocation initiation in the bottom region can be observed too (marked by a black arrow). This image was captured just at the onset of GND initiation within the FCC phase.

\section{Conclusions}

In summary, the current study presents straightforward results on the size and strain rate sensitivity of a eqiumolar (NiCoFeCrGa) high entropy alloy composed of a mixture of BCC/FCC phases. The constituent phases were studied in the mechanical context individually (by deforming pure BCC/FCC pillars), and in case of mixed compositions (pillars with BCC/FCC top-bottom configurations). The current work provides the first clear micromechanical experimental evidence of a transition between thermally activated and drag-controlled deformation mechanisms in FCC and BCC structures. At low strain rates, the thermally activated dislocation motion manifests in weak and monotonous strain rate sensitivity in both BCC and FCC phases. Exceeding the 10$^3$/s strain rate, the drag effect becomes dominant, leading to a sharp increase in the yield stress values. This effect is about 8 times stronger in the BCC phase, that can be attributed to the difference in dislocation characteristics. The measurements are reproducible, and similar tends between experiments and earlier simulation results can be established.

\section*{CRediT author statement} 
\textbf{S. Kalácska:} conceptualization, methodology, validation, formal analysis, investigation, writing -- original draft, writing -- review \& editing, visualization, project administration, funding acquisition, supervision, \textbf{A. Sharma:} validation, investigation, writing - review \& editing, \textbf{R. Ramachandramoorthy:}  investigation, writing -- review \& editing, \textbf{Á. Vida:} writing -- review \& editing, \textbf{F. Tropper:} investigation, \textbf{R. Pero:} resources, \textbf{D. Frey:} resources, \textbf{J. Michler:} funding acquisition, writing -- review \& editing, \textbf{X. Maeder:} funding acquisition, writing -- review \& editing, \textbf{P. D. Ispánovity:} writing -- original draft, writing -- review \& editing, \textbf{G. Kermouche:} formal analysis, funding acquisition, writing -- original draft, writing -- review \& editing.

\section*{Competing interests}  
The authors declare that they have no known competing financial interests or personal relationships that could have appeared to influence the work reported in this paper.

\section*{Acknowledgement}

SK and RR were supported by the EMPA\-POST\-DOCS-II program, which has received funding from the European Union’s Horizon 2020 research and innovation program under the Marie Skłodowska-Curie grant agreement number 754364. SK and GK were funded by the French National Research Agency (ANR) under the project No.~ANR-22-CE08-0012-01 (\emph{INSTINCT}) and No.~ANR-20-CE08-0022 (\textit{RATES}). PDI acknoledges the support by the National Research, Development and Innovation Fund of Hungary under project No.~NKFIH-FK-138975.

\section*{Supporting Information}

Supporting Information and all data are available at the Zenodo repository or from the corresponding author upon reasonable request.

\bibliography{mybib}

\onecolumn
\sloppy
\begin{centering}

\Large{\textbf{Supplementary Material} \textit{for} \\
Micromechanics reveal strain rate dependent transition between dislocation mechanisms in a dual phase high entropy alloy}\\
\vspace{0.5cm}
\normalsize{Szilvia Kal\'{a}cska\emph{$^{a,b,*}$}, Amit Sharma\emph{$^{b}$}, Rajaprakash Ramachandramoorthy\emph{$^{b,c}$}, Ádám Vida\emph{$^{d}$}, Florian Tropper\emph{$^{e}$}, Renato Pero\emph{$^{e}$}, Damian Frey\emph{$^{e}$}, Xavier Maeder\emph{$^{b}$}, Johann Michler\emph{$^{b}$}, Péter Dusán Ispánovity\emph{$^{f}$}, Guillaume Kermouche\emph{$^{a}$}}
\vspace{0.5cm}
\\
\small{\emph{$^{a}$Mines Saint-Etienne, Univ Lyon, CNRS, UMR 5307 LGF, Centre SMS, 158 cours Fauriel 42023 Saint-Étienne, France}\\
\emph{$^{b}$Empa, Swiss Federal Laboratories for Materials Science and Technology, Laboratory of Mechanics of Materials and Nanostructures, CH-3602 Thun, Feuerwerkerstrasse 39. Switzerland}\\
\emph{$^{c}$Max-Planck-Institute für Eisenforschung, Max Planck Strasse 1, 40472 Düsseldorf, Germany}\\
\emph{$^{d}$Dept. of Industrial Materials Technology, Bay Zoltán Nonprofit Ltd. for Applied Research, Kondorfa u.1., H-1116 Budapest, Hungary}\\
\emph{$^{e}$Alemnis AG, Business Park Gwatt, Schorenstrasse 39, 3645 Gwatt (Thun), Switzerland}\\
\emph{$^{f}$ELTE Eötvös Loránd University, Department of Materials Physics, Pázmány Péter sétány 1/a, 1117 Budapest, Hungary}\\}
\vspace{0.5cm}

\end{centering}

\renewcommand{\figurename}{Suppl.~Figure}
\renewcommand{\thefigure}{S\arabic{figure}}
\renewcommand{\tablename}{Suppl.~Table}
\renewcommand{\thetable}{S\arabic{table}}
\setcounter{figure}{0} 
\renewcommand\thesection{S\arabic{section}}
\setcounter{section}{0} 
\renewcommand\bibname{Suppl.~References}

\section{Microstructure characterization}\label{sec:ss0}

\begin{table}[!ht]
    \centering
    \begin{tabular}{|l|l|l|l|l|l|l|}
    \hline
        Sample & Phase & GAD & Mean & Min. & Max. & SD \\ 
        name & ID & ($\mu$m) & ($\mu$m) & ($\mu$m) & ($\mu$m) & ($\mu$m) \\ \hline
        S1 & FCC & 18.64 & 45.27 & 5.35 & 97.92 & 14.41 \\ 
        S1 & BCC & 25.52 & 107.66 & 5.35 & 174.72 & 30.9 \\ \hline
        S2 & FCC & 87.92 & 476.19 & 10 & 2127 & 184.95 \\ 
        S2 & BCC & 208 & 1658.86 & 10 & 5610 & 550.69 \\ \hline
    \end{tabular}
    \caption{Grain statistics from large area EBSD maps, determined using AZtecCrystal v2.2. GAD: grain average diameter based on equivalent circle diameter, SD: standard deviation.}\label{table:s0}
\end{table}

\begin{figure}[!ht]
    \centering
    \includegraphics[width=0.49\textwidth]{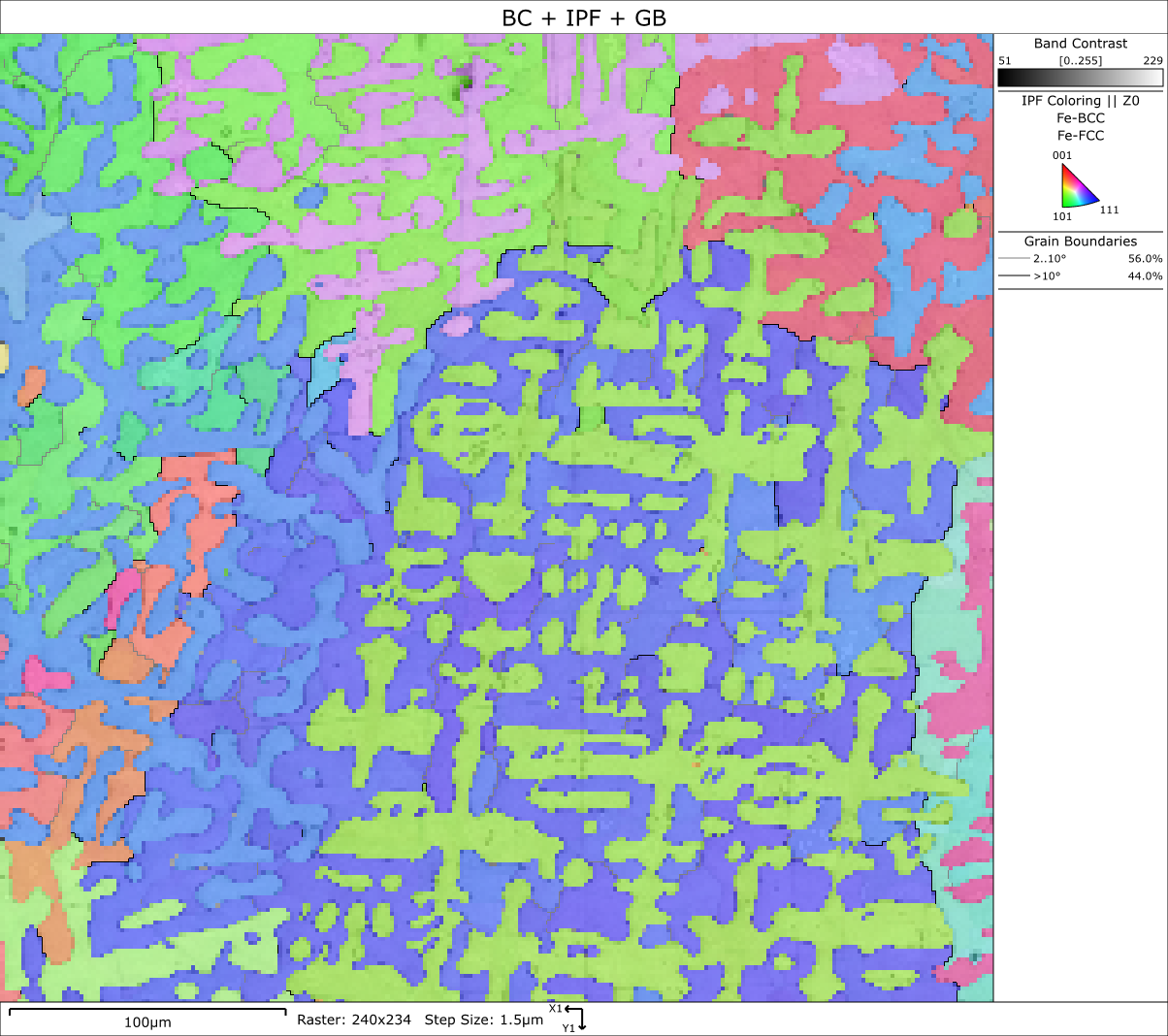}
    \includegraphics[width=0.49\textwidth]{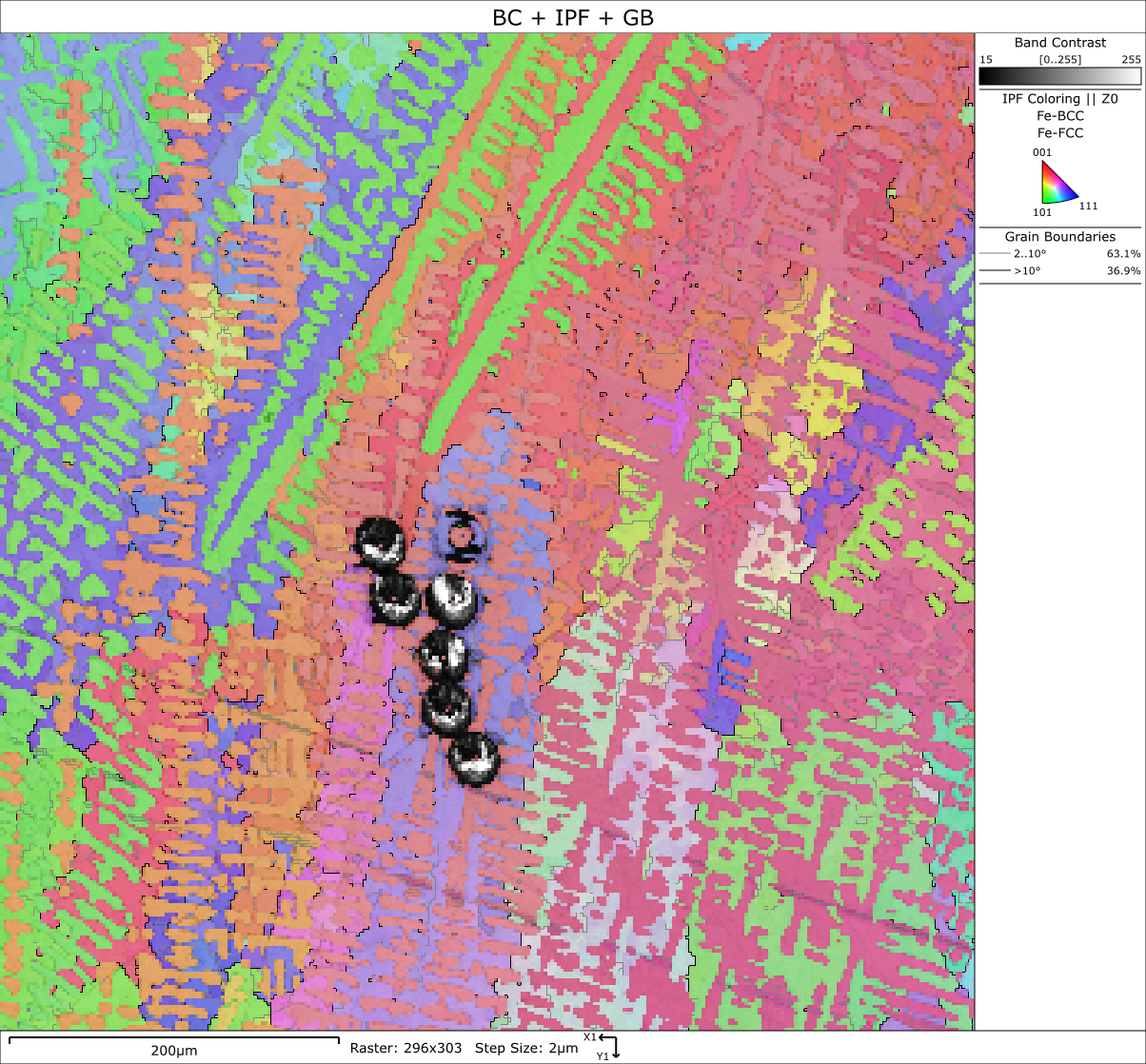}
    \caption{Large area EBSD maps for grain statistics. (left) Sample 1, (right) Sample 2 mapped with the first pillars. BC: band contrast, IPF: inverse pole figure in Z direction, GB: grain boundaries.
    \label{fig:s1}}
\end{figure}

\clearpage
\section{Sample preparation}\label{sec:ss1}

\noindent

\begin{figure}[!ht]
    \centering
    \includegraphics[width=0.45\textwidth]{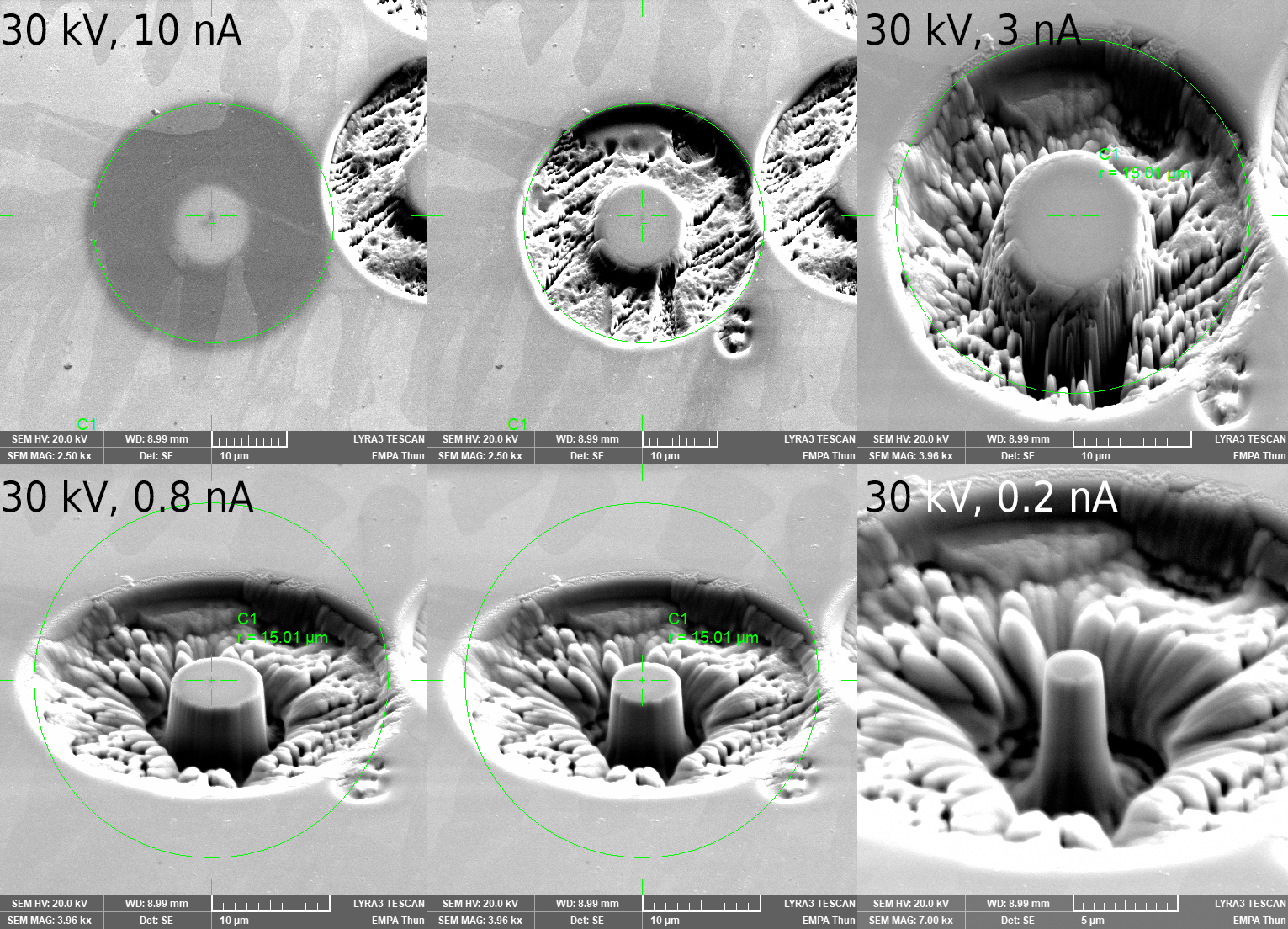}
    \includegraphics[width=0.45\textwidth]{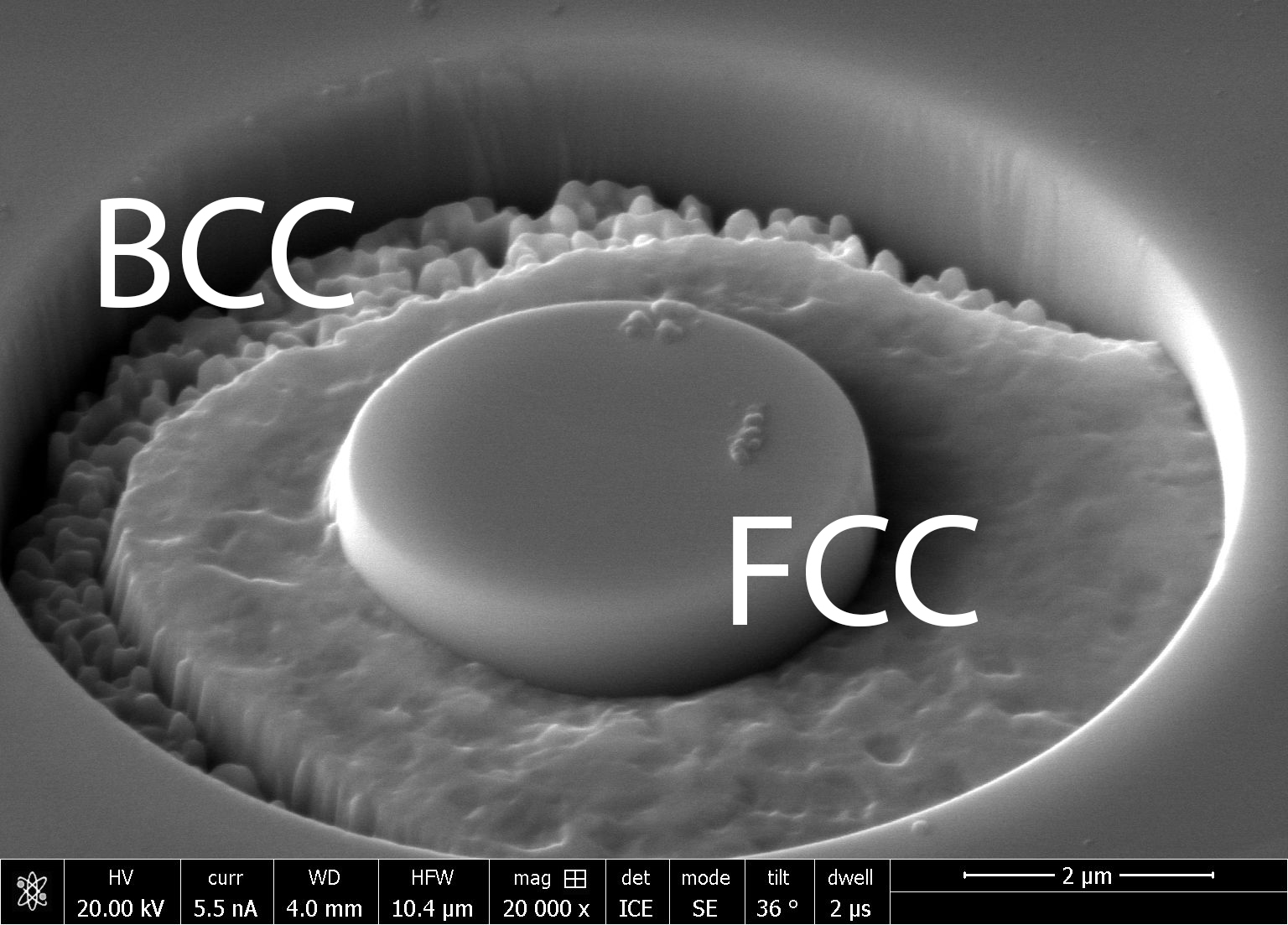}
    \caption{(left) \textbf{Pillar fabrication, Sample 1.} (right) \textbf{Difference in sputtering speed between BCC and FCC phases.} SE image was take during pillar preparation (F7). The inner part is FCC, and the outer part of the crater overlaps with a BCC boundary. Image was taken at a 36$^{\circ}$ tilt with a 16$^{\circ}$ pre-tilted holder).
    \label{fig:s2}}
\end{figure}

\begin{figure}[!ht]
    \centering
    \includegraphics[width=0.99\textwidth]{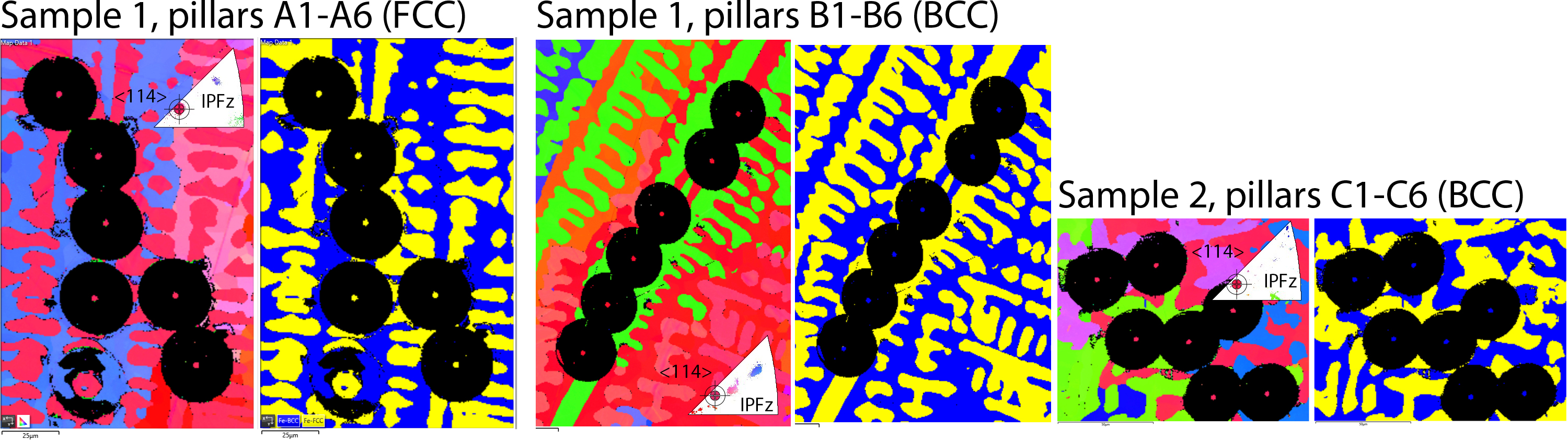}
    \caption{\textbf{Pillars, series A, B, C}, Sample 1. IPF and phase maps of the three kinds of pillars prepared. Standard IPF triangles show only the pixels belonging to the pillar's phase. Phase map colors: FCC -- yellow, BCC -- blue.
    \label{fig:s3}}
\end{figure}

\begin{figure}[!ht]
    \centering
    \includegraphics[width=0.4\textwidth]{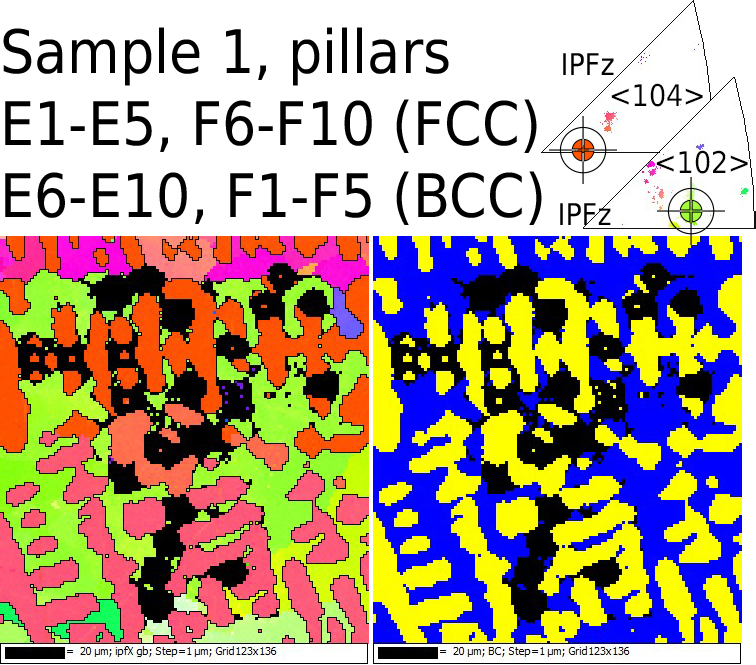}
    \hspace{0.5cm}
    \includegraphics[width=0.55\textwidth]{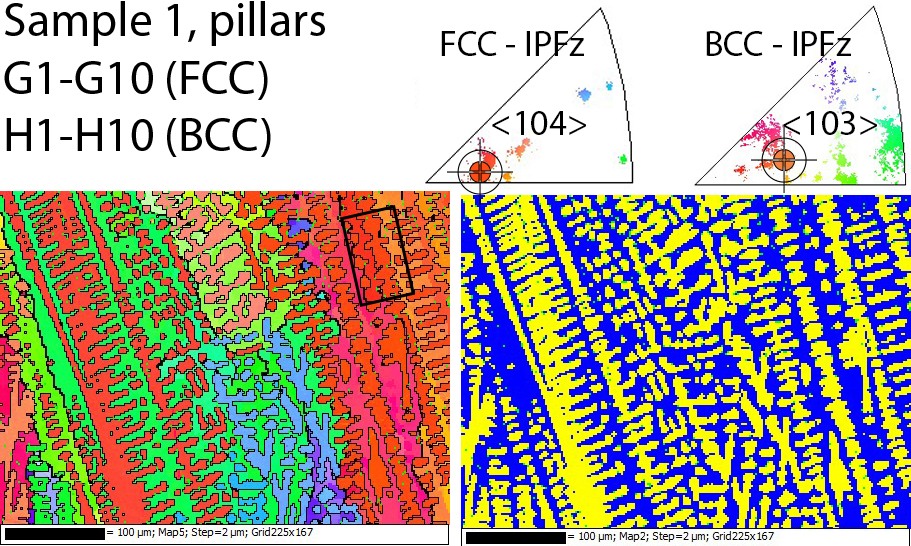}
    \caption{\textbf{Pillars, series E, F, G, H}, Sample 1. IPF and phase maps of the two kinds of pillars prepared. Black square marks the final position of the pillar series G and H. Phase map colors: FCC -- yellow, BCC -- blue.
    \label{fig:s4}}
\end{figure}

\begin{figure}[!ht]
    \centering
    \includegraphics[width=0.50\textwidth]{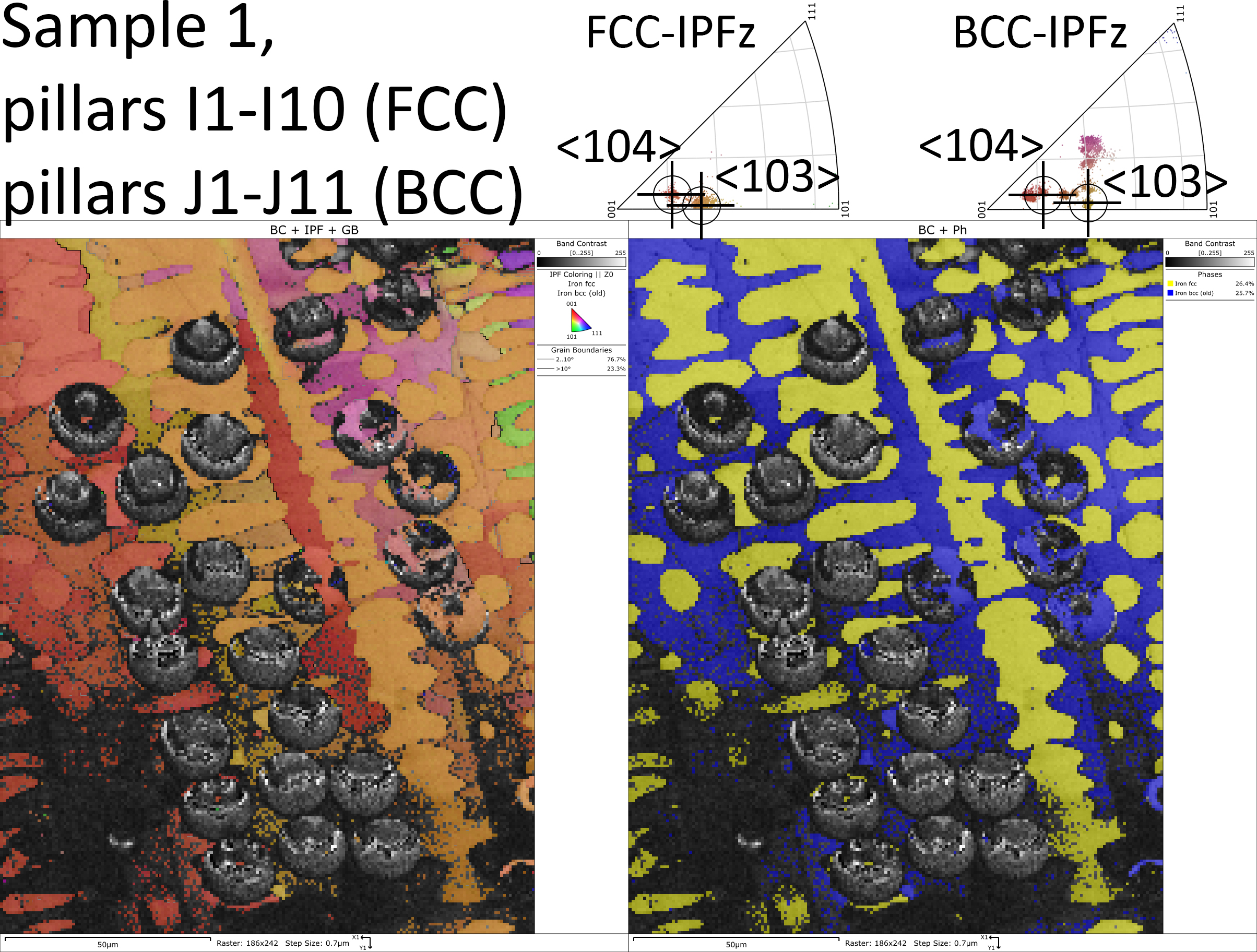}
    \caption{\textbf{Pillars, series I and J}, Sample 1. IPF and phase maps of the BCC pillars prepared. Phase map colors: FCC -- yellow, BCC -- blue.
    \label{fig:s5}}
\end{figure}

\begin{figure}[!ht]
    \centering
    \includegraphics[width=0.45\textwidth]{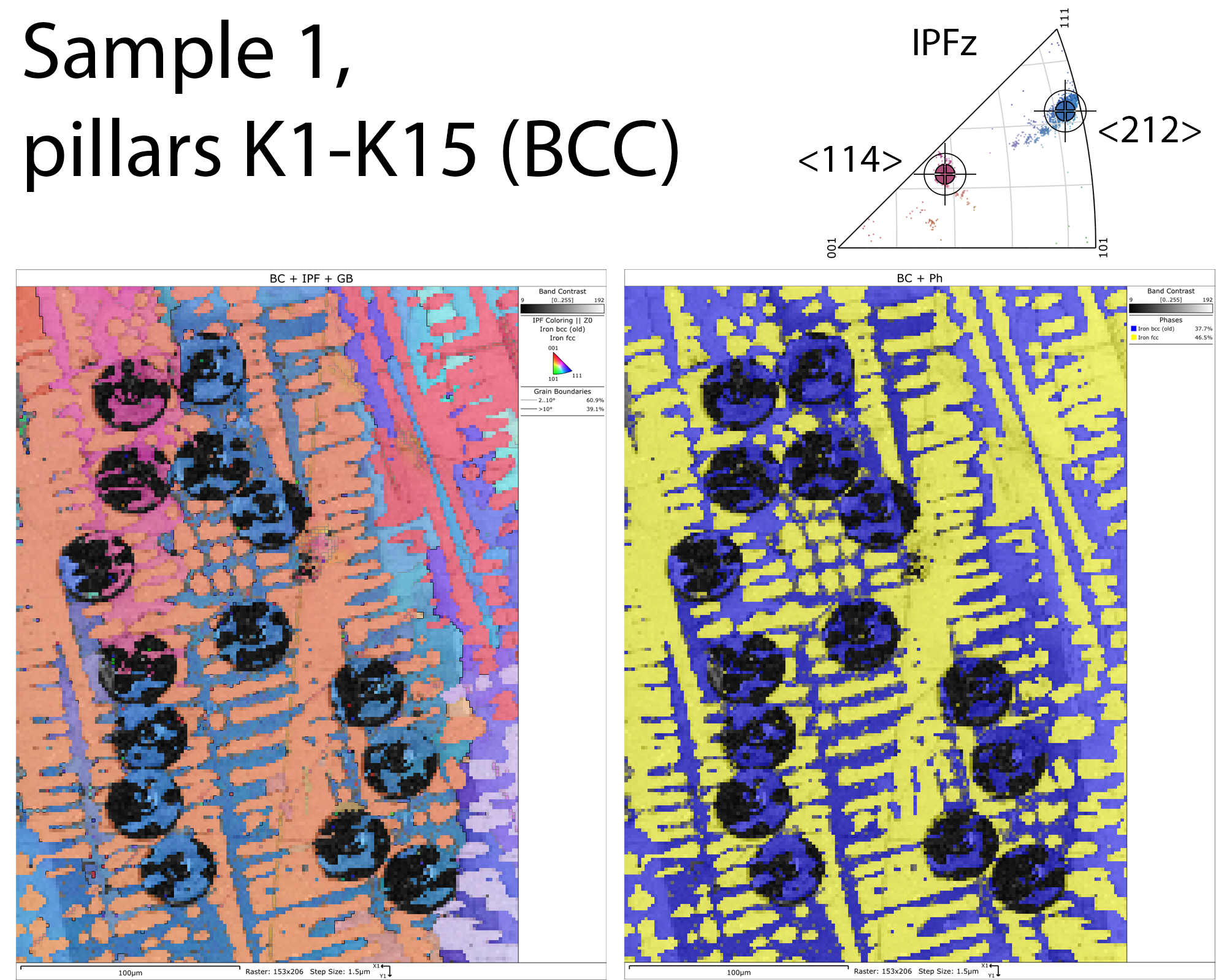}
    \hspace{0.2cm}
    \includegraphics[width=0.45\textwidth]{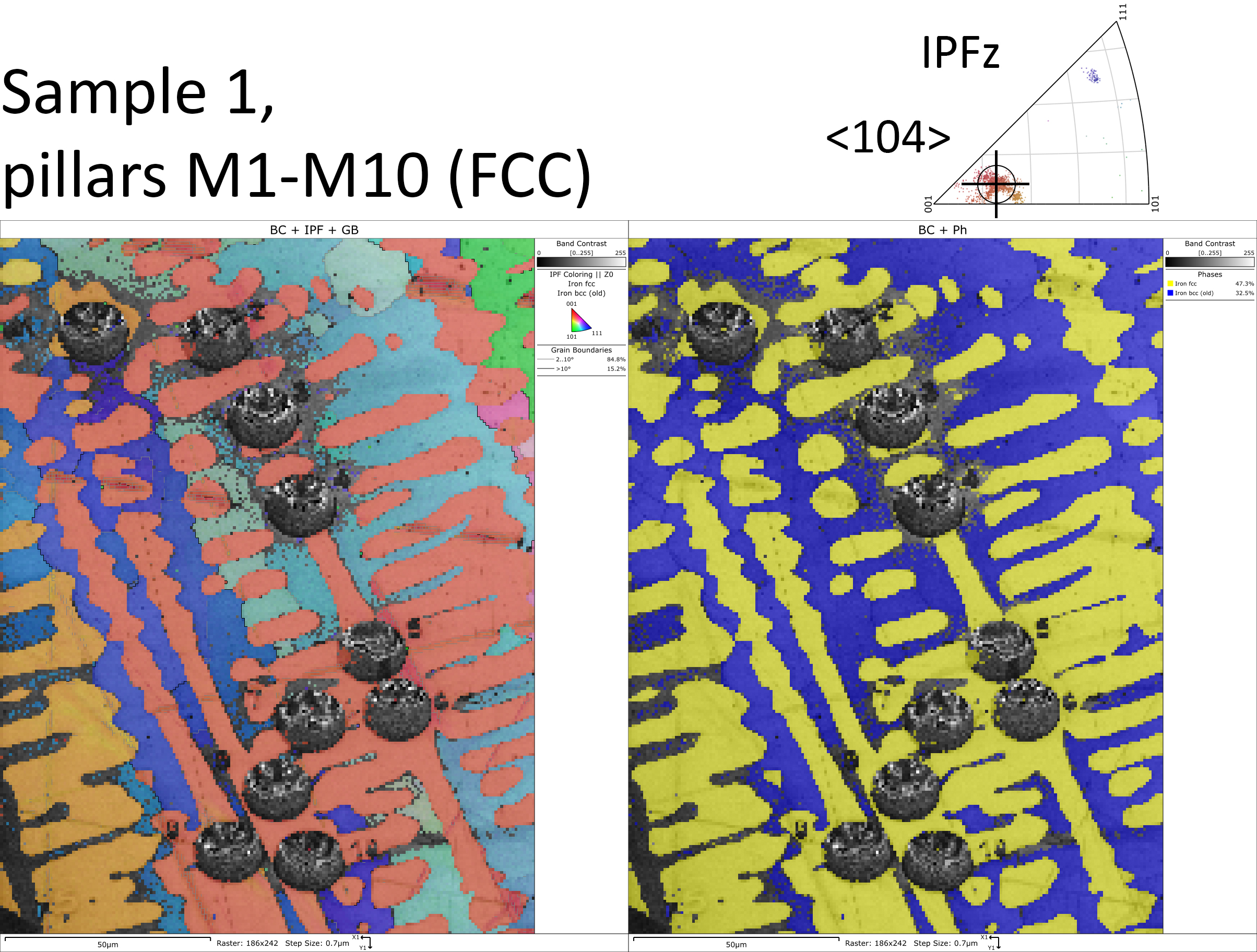}
    \caption{\textbf{Pillars, series K and M}, Sample 1. IPF and phase maps of the BCC pillars prepared. Phase map colors: FCC -- yellow, BCC -- blue.
    \label{fig:s6}}
\end{figure}

\begin{figure*}[!ht]
    \centering
    \includegraphics[width=0.70\textwidth]{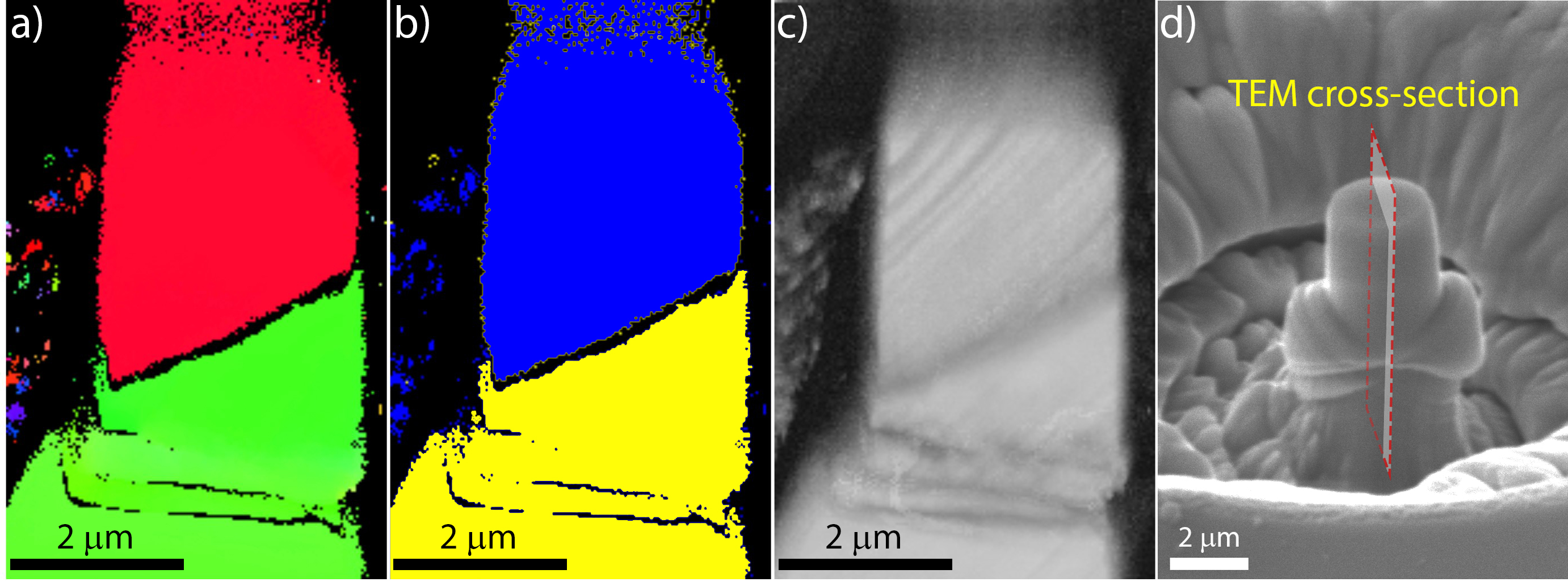}
    \caption{\textbf{EBSD results}, Sample 1, pillar B3. (a) IPF-Y orientation map (coordinate-system origin is at the bottom left corner: X pointing towards the right, Y pointing upwards).  (b) Phase map with colors: FCC -- yellow, BCC -- blue. (c) Band contrast image. FIB milling curtaining effect (with $\sim 42^{\circ}$ angle w.r.t. the horizontal direction) can be observed. (d) TEM cross-section plane sketched over the deformed pillar.
    \label{fig:s7}}
\end{figure*}

\begin{figure*}[!ht]
    \centering
    \includegraphics[width=0.2\textwidth]{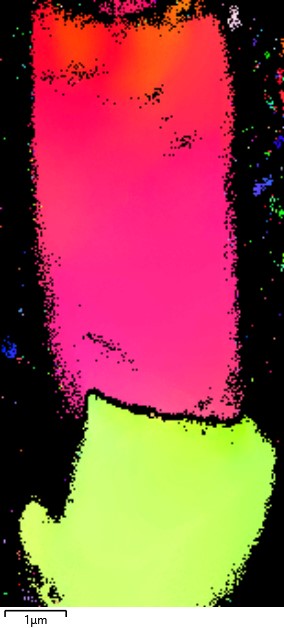}
    \includegraphics[width=0.2\textwidth]{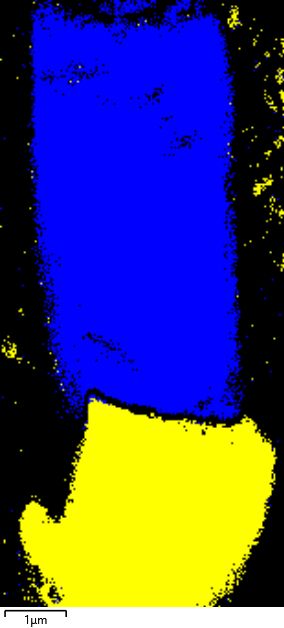}
    \includegraphics[width=0.2\textwidth]{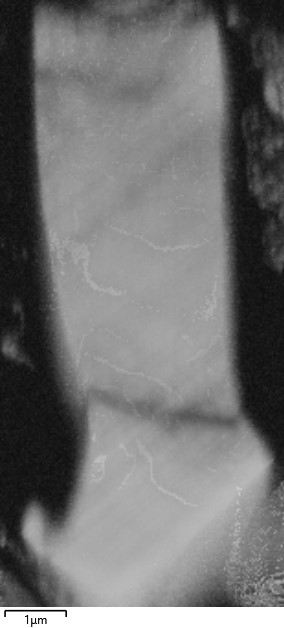}
    \caption{\textbf{EBSD results}, Sample 1, pillar B6. (left) IPF-Y orientation map. (middle) phase map with colors: FCC -- yellow, BCC -- blue. (right) Band contrast image.
    \label{fig:s8}}
\end{figure*}

\section{Poisson's ratio}\label{sec:ss2}

Two ways of calculating the Poisson's ratio have been considered in this work. The first is to simply average the Poisson's ratio of the constituent elements:

\begin{equation}
    \nu_1=\frac{1}{5} \sum  \nu_{\mathrm{{i}}} = 0.321, \quad \mathrm{i}: \mathrm{Ni, Co, Fe, Cr, Ga}
\end{equation}

\noindent with $\nu_{Ni}=0.31$, $\nu_{Co}=0.32$, $\nu_{Fe}=0.293$, $\nu_{Cr}=0.21$, $\nu_{Ga}=0.47$. These values were taken from the Goodfellow catalogue of high purity materials \cite{goodfellow}.

The second method is performed by measuring ultrasonic velocities in the material, and calculating $\nu_2 = 0.294$ by the formula

\begin{equation}
    \nu_2 = \frac{v_p^2 - 2 v_s^2}{2(v_p^2 - v_s^2)},
\end{equation}

\noindent where $v_p = 5.22 \times 10^3$ m/s longitudinal and $v_p = 2.82 \times 10^3$ m/s transversal velocities were measured earlier. Eventually, the average of the two calculated values were used in the evaluations of the current work:

\begin{equation}
    \overline{\nu} = \frac{1}{2}(\nu_1 + \nu_2) \cong 0.31.
\end{equation}

\noindent The uncertainty of the Poisson's ratio results in $\sim $1\% error in the Young's modulus values.

\clearpage
\section{Micromechanical details}

\begin{figure}[!ht]
    \centering
    \includegraphics[width=0.99\textwidth]{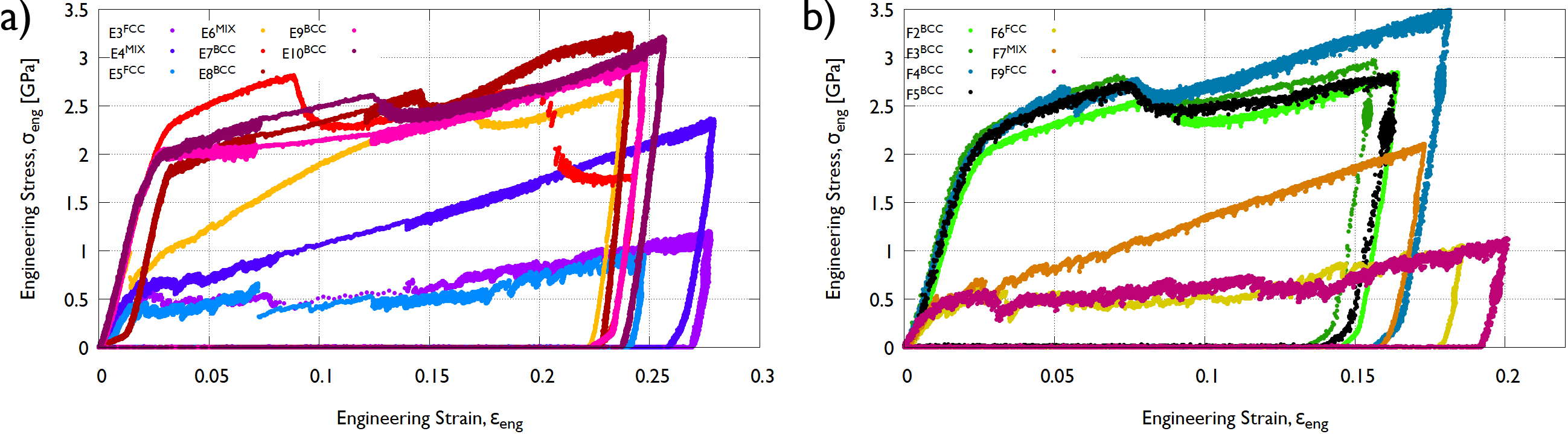}
    \caption{\textbf{Pillar deformation, series E, F.} Sample 1. Stress-strain curves at quasi-static strain rates. Pillars with (a) 2.5 $\upmu$m and (b) 1.5 $\upmu$m diameters. Phase map colors: FCC -- yellow, BCC -- blue. The $\sigma_{eng} - \varepsilon_{eng}$ curve of pillar ``E7" exhibits a slightly different behaviour at the beginning of the loading ($\varepsilon <0.01$), this is due to the fact that the surface of the pillar was not parallel with the flat punch tip. A unique misalignment like this (when preparing a batch of pillars) is most likely produced by imperfect FIB milling, where a slight beam misalignment would unevenly expose an area, causing the halo of the ion beam to touch only one part of the sample.
    \label{fig:s9}}
\end{figure}

\begin{figure}[!ht]
    \centering
    \includegraphics[width=0.6\textwidth]{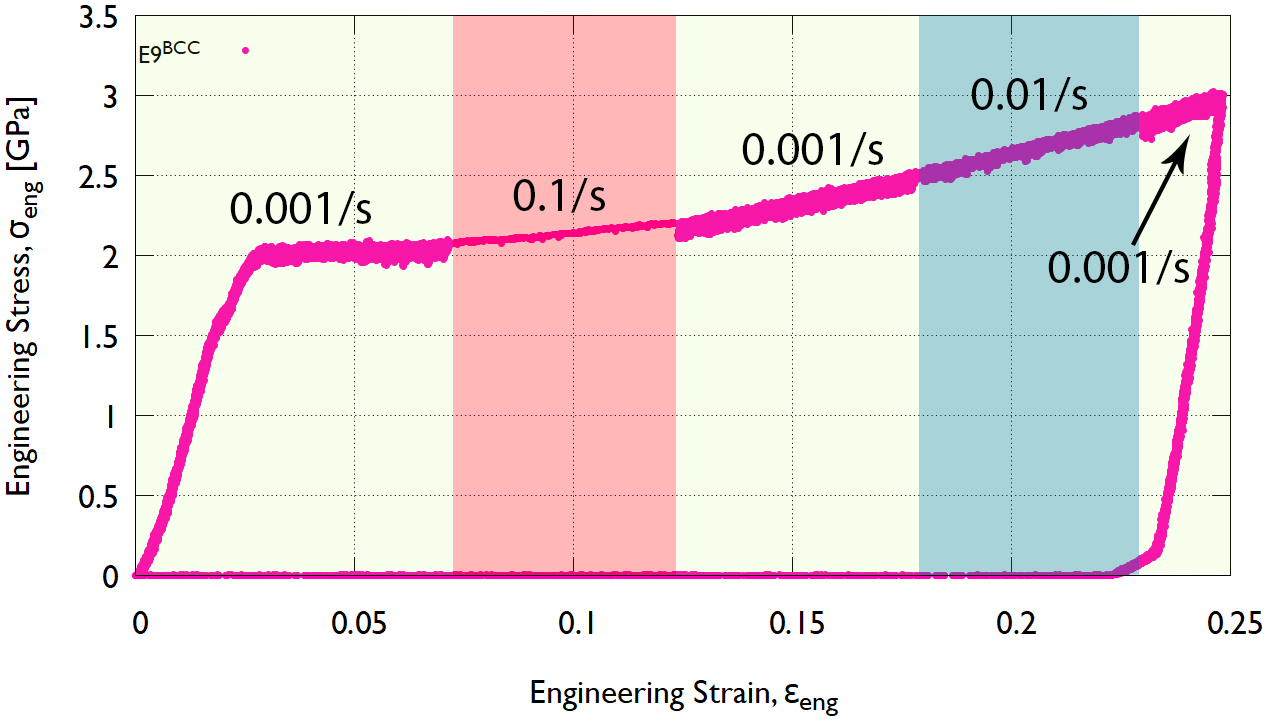}
    \caption{\textbf{Strain rate jump test}, Sample 1, over the example of the E9 $\sigma_{eng}-\varepsilon_{eng}$ curve.
    \label{fig:s10}}
\end{figure}

\begin{figure}[!ht]
    \centering
    \includegraphics[width=0.99\textwidth]{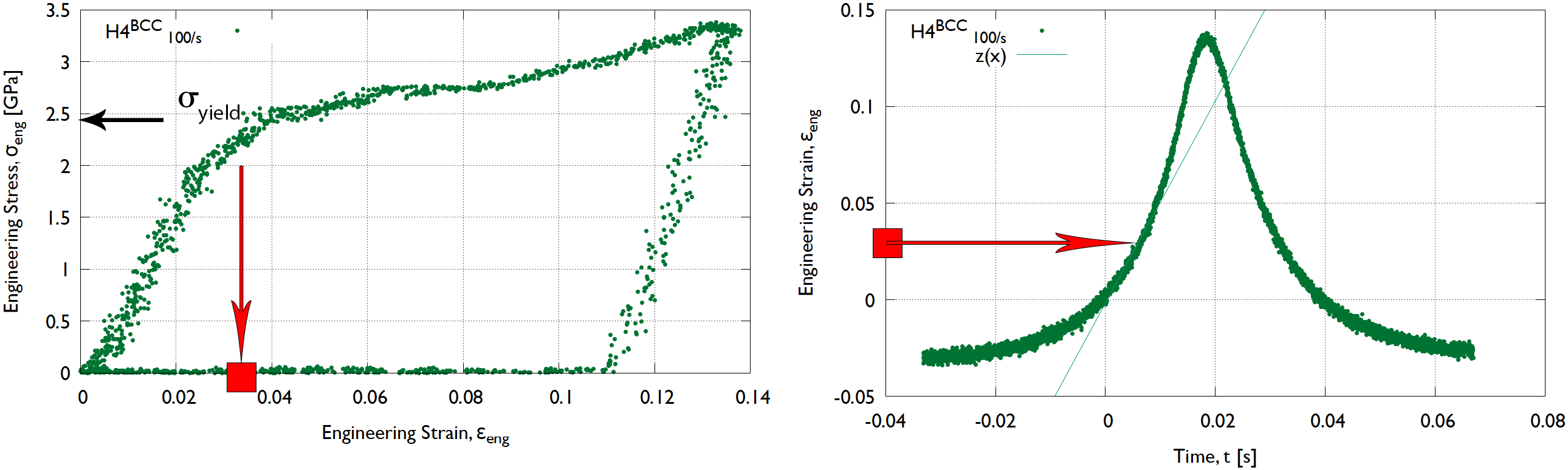}
    \caption{\textbf{Proportional loading profile.} Sample 1. $\varepsilon_{eng}$ variation over time. At the $\varepsilon_{eng}$ position related to the $\sigma_{yield}$, a ``z(x)" linear fit was performed to determine the strain rate upon yielding.
    \label{fig:s11}}
\end{figure}

\begin{figure}[!ht]
    \centering
    \includegraphics[width=0.5\textwidth]{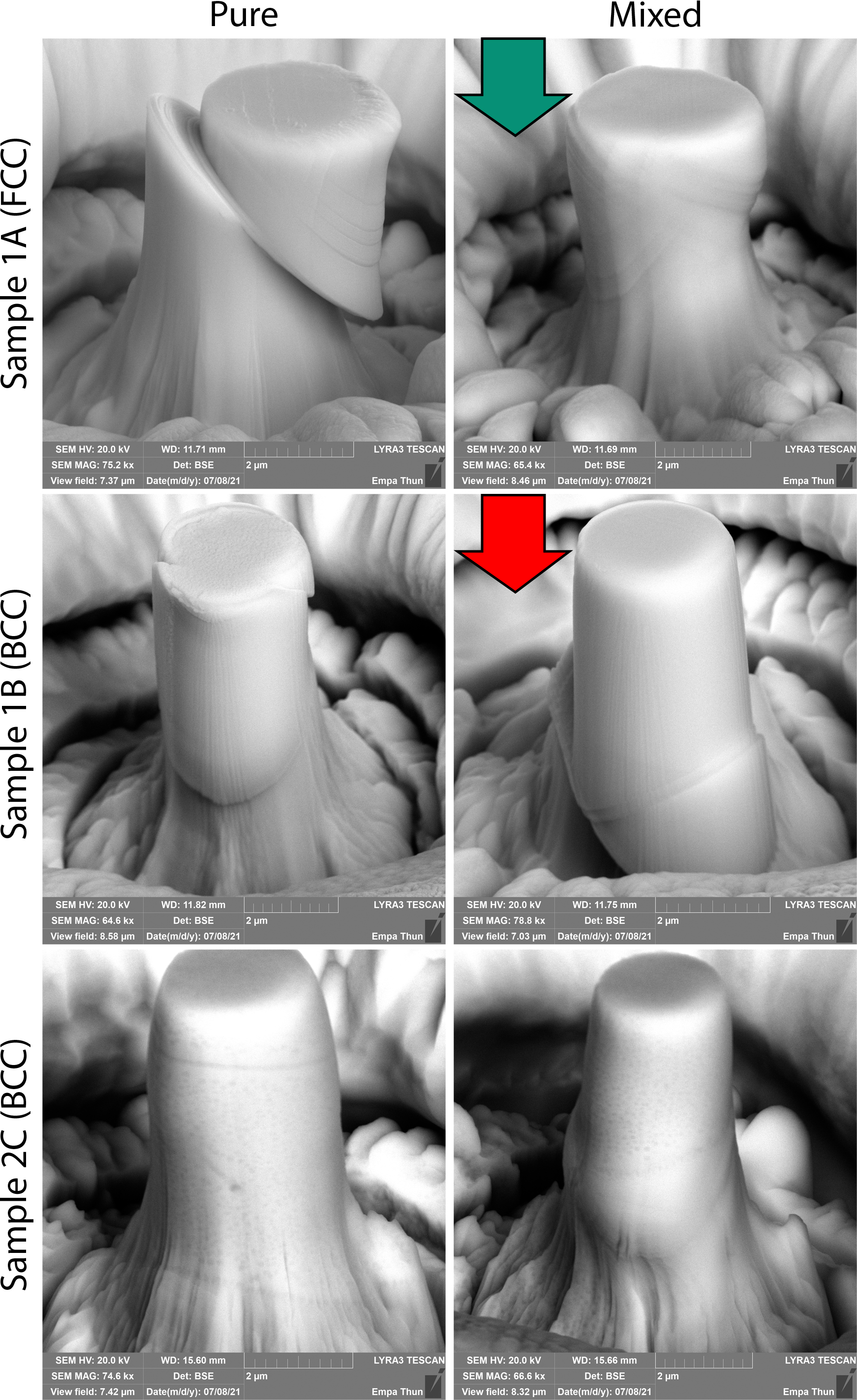}
    \caption{\textbf{Deformation modes} BSE images
    \label{fig:s12}}
\end{figure}

\begin{figure}[!ht]
    \centering
    \includegraphics[width=0.6\textwidth]{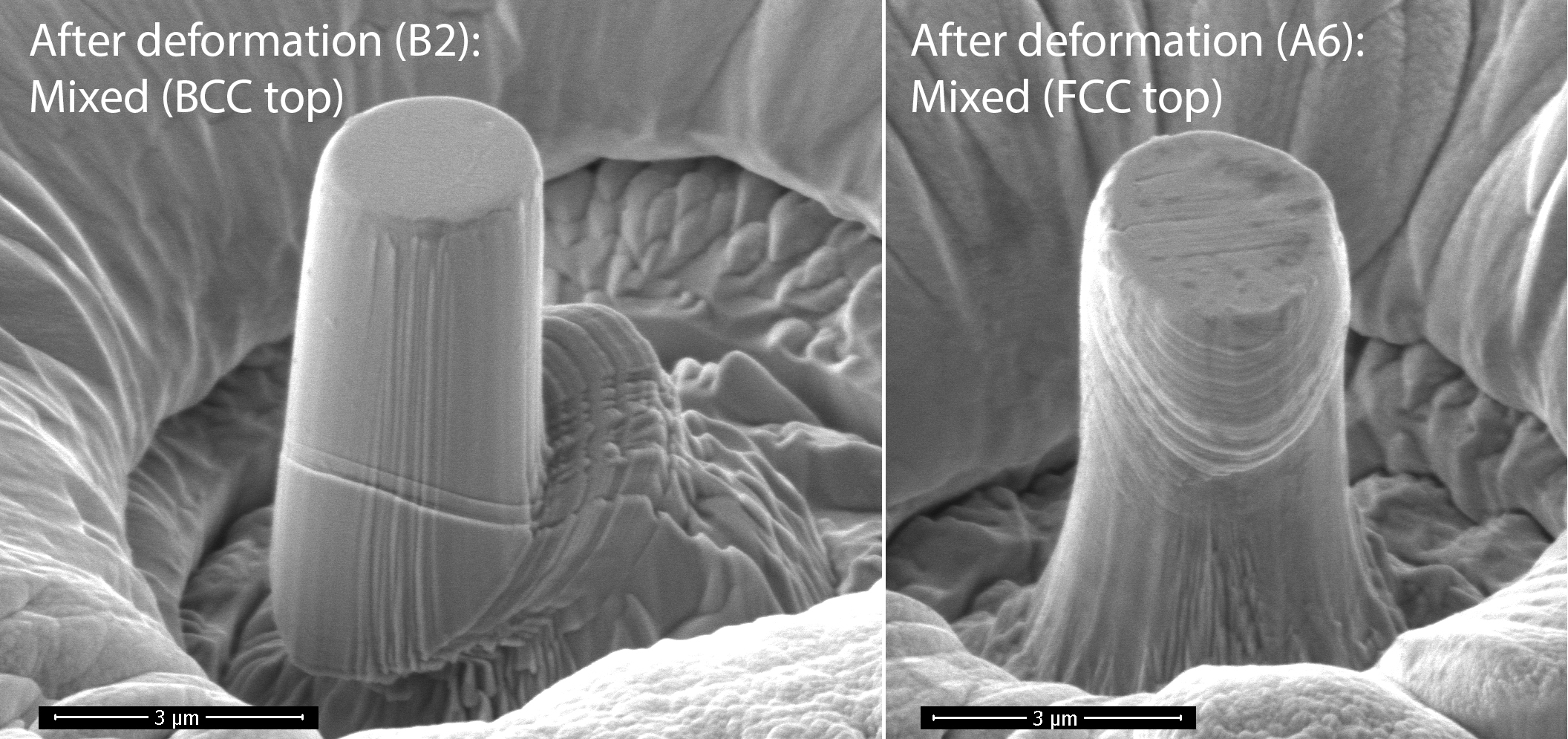}
    \caption{\textbf{Mixed deformation modes}, Sample 1, SE images.
    \label{fig:s13}}
\end{figure}

\clearpage
\section{\textit{Post mortem} microstructure characterization}

\begin{figure}[!ht]
    \centering
    \includegraphics[width=0.60\textwidth]{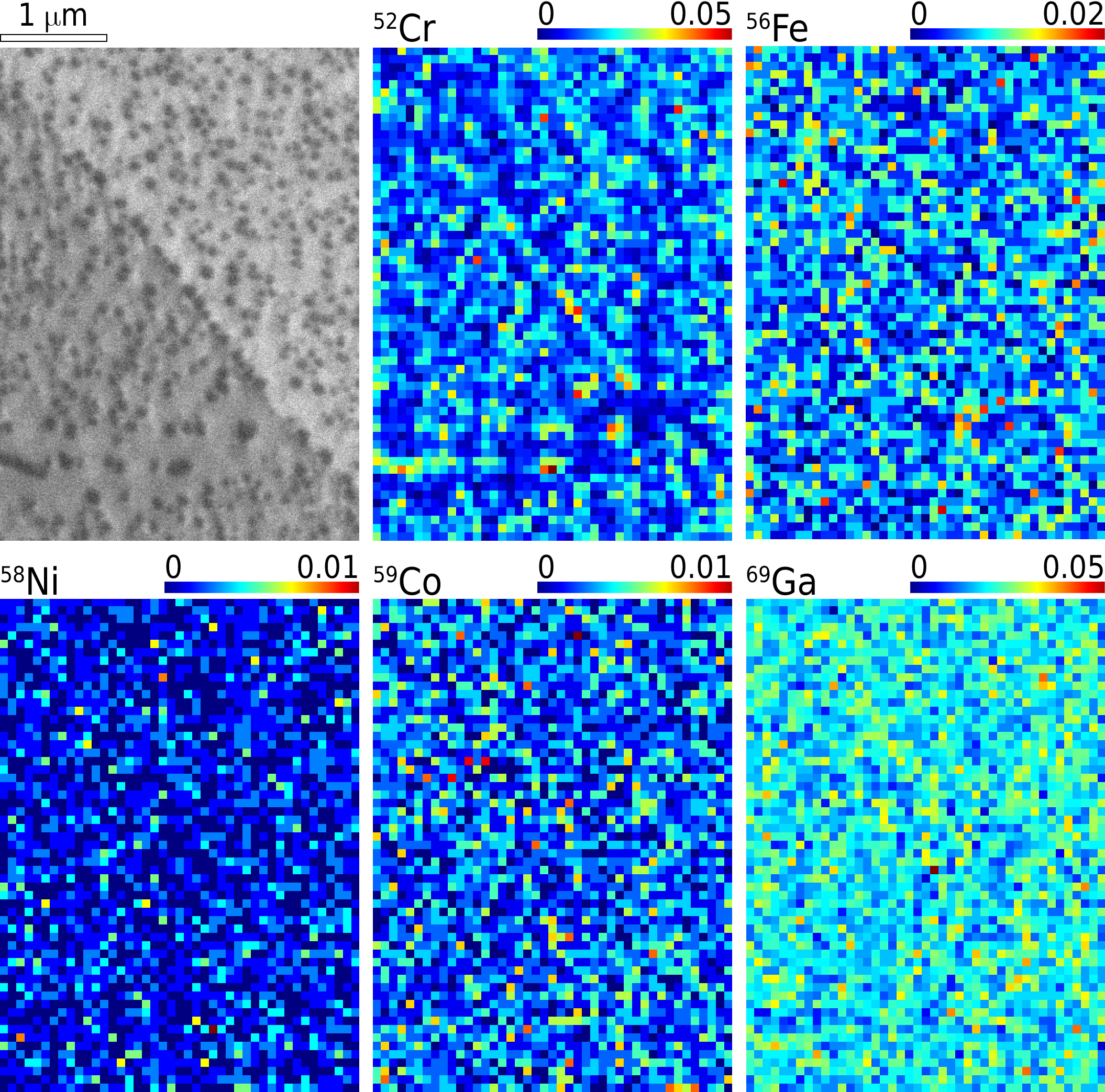}
    \caption{\textbf{TOF-SIMS} results along a BCC grain boundary, Sample 2. SE image (in greyscale) highlights the grain boundary  between the brighter/darker regions. Many precipitates align with the grain boundary. Coloured images show distributions of the marked ions. Color scales are in arbitrary units (counts/TOF extraction).
    \label{fig:s14}}
\end{figure}

\begin{figure*}[!ht]
    \centering
    \includegraphics[width=0.75\textwidth]{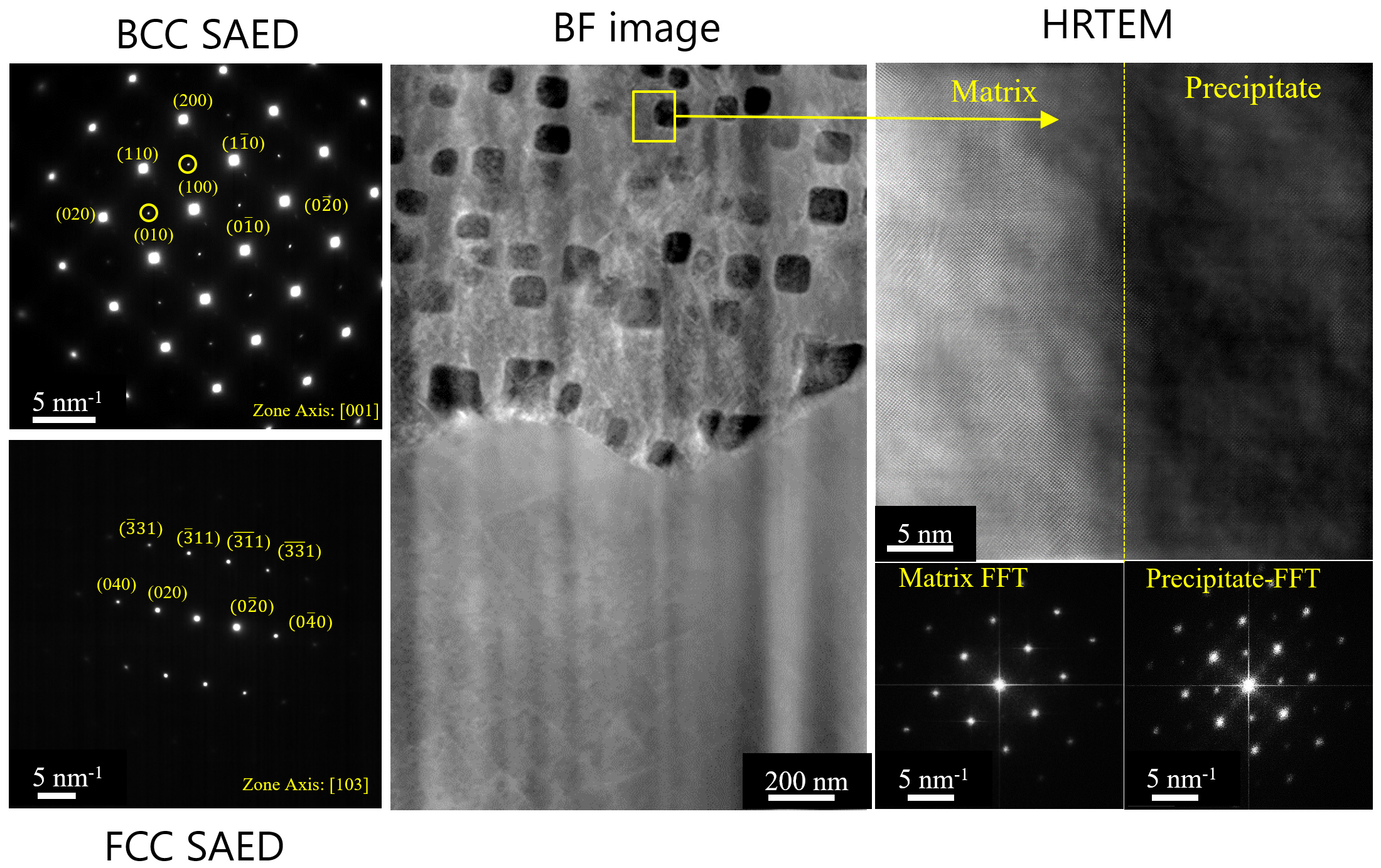}
    \caption{\textbf{TEM imaging} pillar C1, Sample 2 at the BCC (containing precipitates) and FCC interface.
    \label{fig:s15}}
\end{figure*}

\begin{figure*}[!ht]
    \centering
    \includegraphics[width=0.99\textwidth]{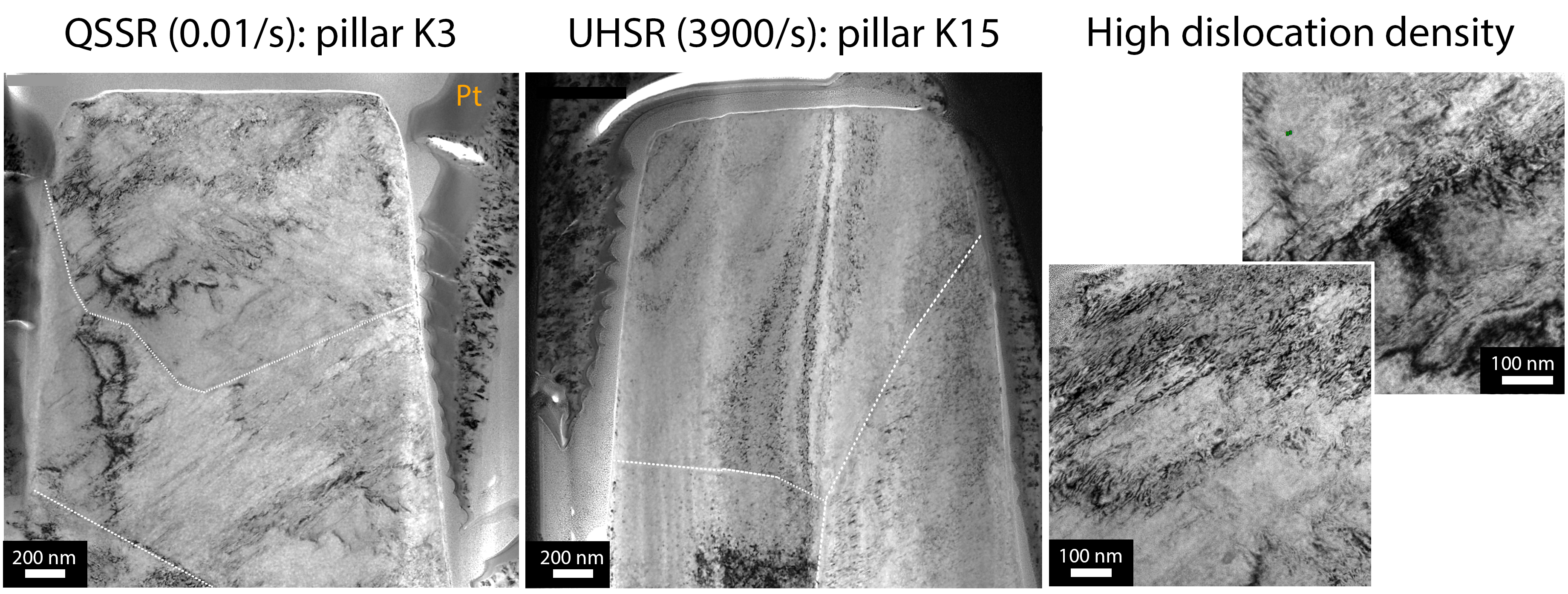}
    \caption{\textbf{TEM imaging} Sample 1, pillars at various strain rates. Grain boundaries are marked with dotted white lines. \label{fig:s16}}
\end{figure*}

\begin{figure*}[!ht]
    \centering
    \includegraphics[width=0.95\textwidth]{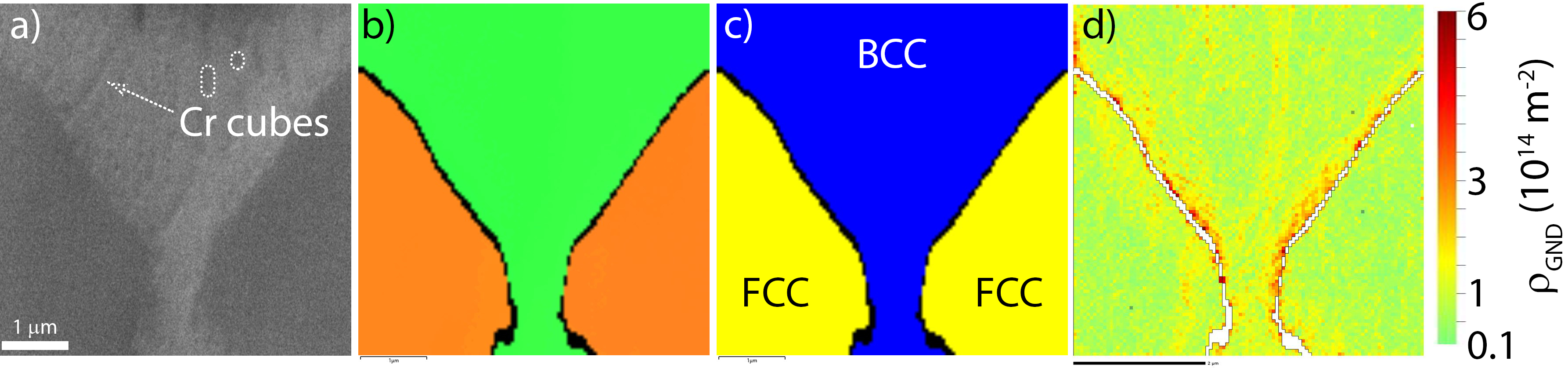}
    \caption{\textbf{HR-EBSD}, Sample 2, undeformed specimen. (a) Band contrast image of the mapped area. A few precipitates are highlighted with white dashed shapes. (b) IPFz orientation map. (c) Phase contrast map. (d) Map of the estimated GND-density.
    \label{fig:s17}}
\end{figure*}

\begin{figure*}[!ht]
    \centering
    \includegraphics[width=0.99\textwidth]{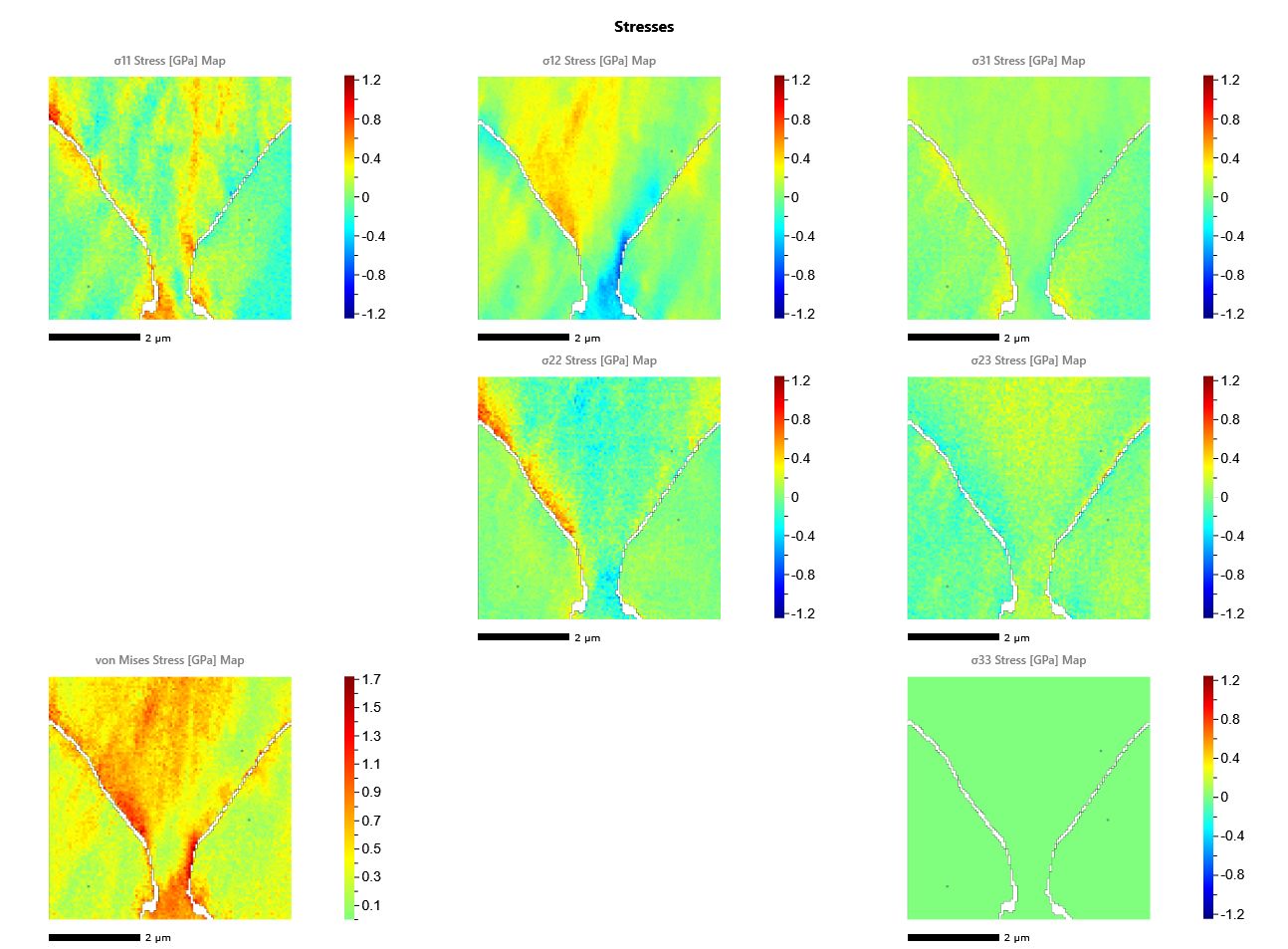}
    \caption{\textbf{HR-EBSD stress tensor components ($\sigma_{ij}$) and von Mises stress}. Sample 2, undeformed specimen. 
    \label{fig:s18}}
\end{figure*}

\begin{figure}[!ht]
    \centering
    \includegraphics[width=0.4\textwidth]{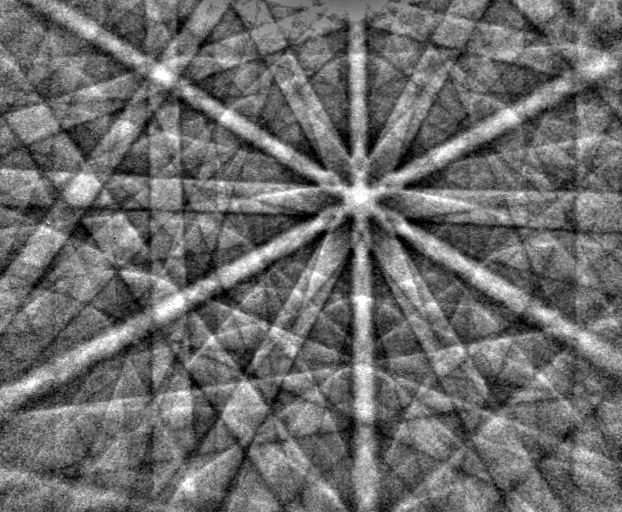}
    \caption{\textbf{Typical EBSD pattern}, Sample 2.
    \label{fig:s19}}
\end{figure}

\begin{figure}[!ht]
    \centering
    \includegraphics[width=0.8\textwidth]{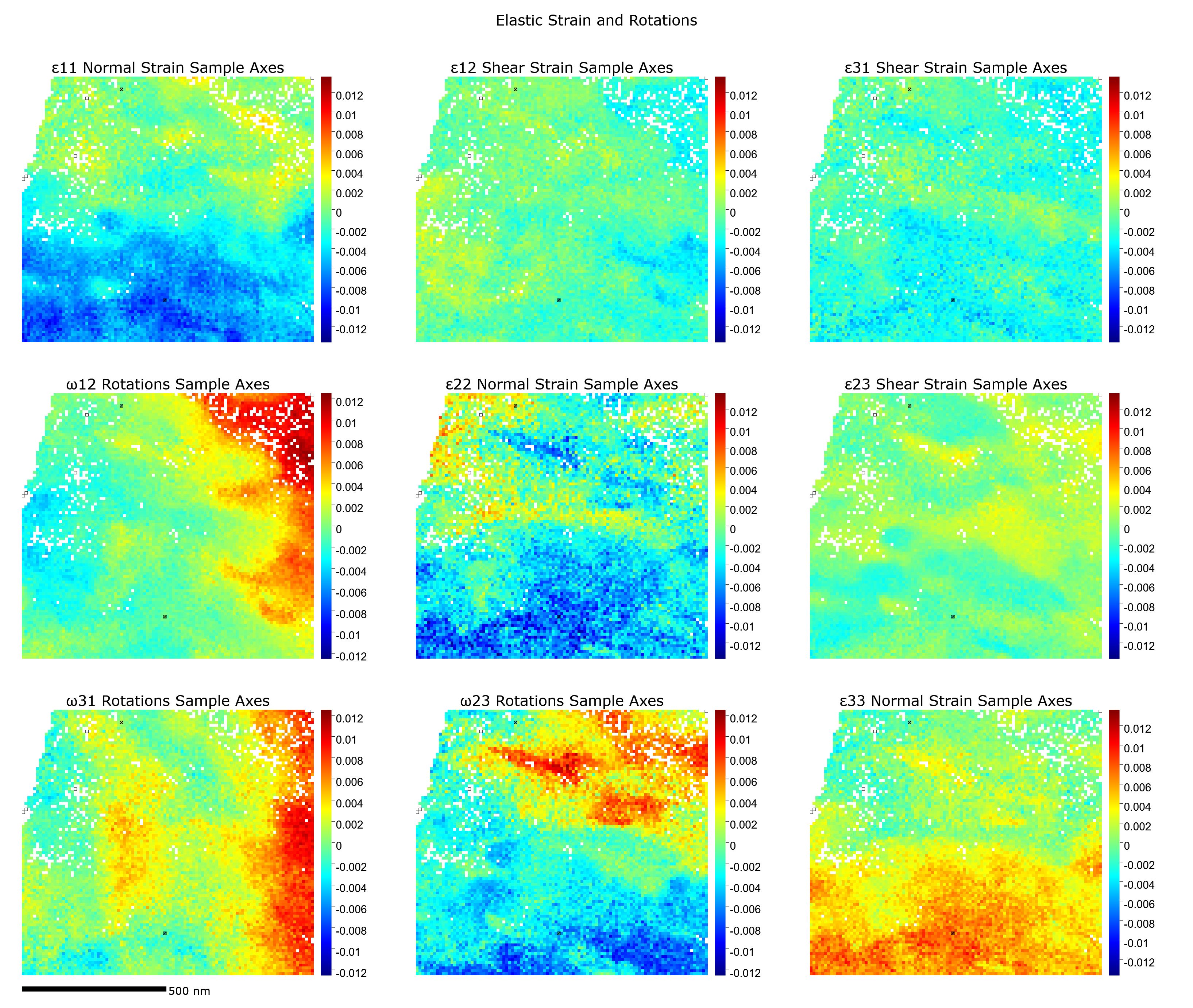}
    \caption{\textbf{HR-TKD}, C1 strain- and rotation tensor elements, BCC phase.
    \label{fig:s20}}
\end{figure}

\begin{figure}[!ht]
    \centering
    \includegraphics[width=0.75\textwidth]{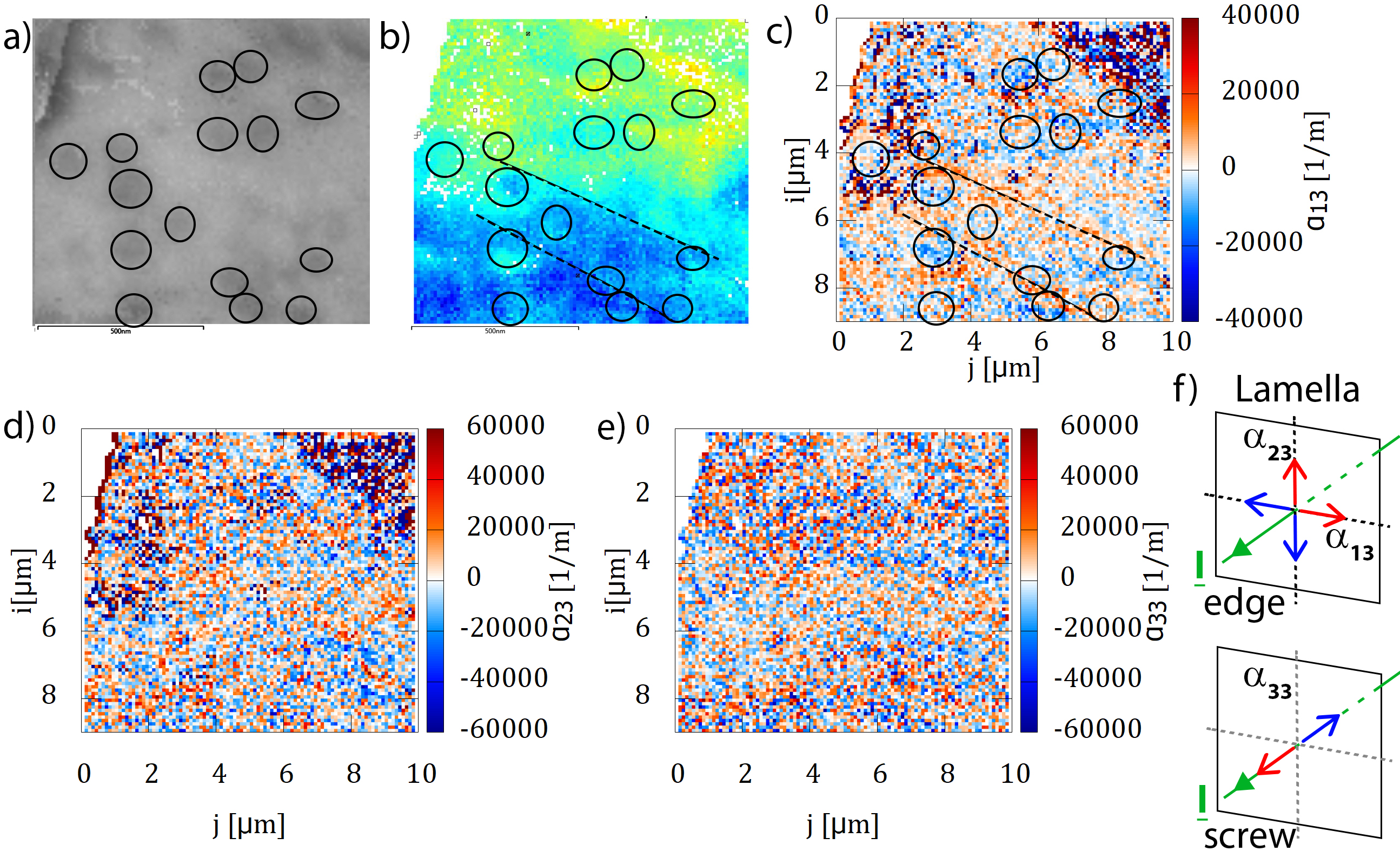}
    \caption{\textbf{HR-TKD}, C1 Burgers vector analysis on the BCC region. a) Band contrast image, b) $\varepsilon_{11}$ map (colors identical to Fig. \ref{fig:s20}) and c) $\alpha_{13}$ map comparison. Black circles indicate the presence of precipitates, dotted lines show edge type GND density features similar to Fig. \ref{fig:05}. d) $\alpha_{23}$ and e) $\alpha_{33}$ maps. f) Sketch  indicating the typical Burgers vector directions based on the component analysis.
    \label{fig:s21}}
\end{figure}

\clearpage
\section{Pillars' Summary}\label{sec:sspillars}

\begin{table}[!ht]
    \centering
    \begin{tabular}{|l|l|l|l|l|l|l|l|p{0.3in}|l|l|l|l|p{0.5in}|}
    \hline
        Pillar& ID & Phase & Avg. & Avg. & Height & Def. & Max. & Tip & SR & Yield & Smp & Load & Notes \\ 
        name & & & diam. & height & (TC)& mode & def.& $v$ & [/s] & stress & [Hz] & cell & \\ 
          & & & [$\mu$m] & [$\mu$m] & [$\mu$m] & & (\%) &  [nm/s] &  & [MPa] & & & \\ \hline
        \textit{A1} & S1 & FCC & 3.31 & 5.28 & 6.45 & pure & 23 & 8 & 0.001 & 273 & 40 & SLC-1 & ~ \\ 
        \textit{A2} & ~ & ~ & 3.39 & 5.16 & 6.30 & pure & 17 & ~ & 0.001 & 197 & 40 & ~ & ~ \\ 
        \textit{A3} & ~ & ~ & 3.26 & 5.52 & 6.74 & mixed & 16 & ~ & 0.001 & 156 & 40 & ~ & LO/TEM \\ 
        \textit{A4} & ~ & ~ & 3.33 & 5.64 & 6.89 & pure & 17 & ~ & 0.001 & 196 & 40 & ~ & ~ \\ 
        \textit{A5} & ~ & ~ & 3.35 & 6.32 & 7.72 & pure & 16 & ~ & 0.001 & 211 & 40 & ~ & ~ \\ 
        \textit{A6} & ~ & ~ & 3.33 & 5.28 & 6.45 & mixed & 17 & ~ & 0.001 & 166 & 40 & ~ & ~ \\ \hline
        \textit{B1} & S1 & BCC & 3.15 & 5.91 & 7.21 & mixed & 17 & 8 & 0.001 & 876 & 40 & SLC-1 & ~ \\ 
        \textit{B2} & ~ & ~ & 3.14 & 5.91 & 7.21 & mixed & 17 & ~ & 0.001 & 982 & 40 & ~ & ~ \\ 
        \textit{B3} & ~ & ~ & 3.22 & 5.72 & 6.98 & mixed & 17 & ~ & 0.001 & 857 & 40 & ~ & LO/HR-EBSD \\
        \textit{B4} & ~ & ~ & 3.42 & 6.08 & 7.42 & mixed & 17 & ~ & 0.001 & 643 & 40 & ~ & ~ \\ 
        \textit{B5} & ~ & ~ & 3.29 & 6 & 7.32 & pure & 17 & ~ & 0.001 & 1828 & 40 & ~ & ~ \\ 
        \textit{B6} & ~ & ~ & 3.27 & 5.88 & 7.18 & pure & 15 & ~ & 0.001 & 1845 & 40 & ~ & LO/HR-EBSD \\ \hline
        \textit{C1} & S2 & BCC & 3.22 & 5.44 & 6.64 & pure & 16 & 8 & 0.001 & 1035 & 40 & SLC-1 & LO/TEM \\
        \textit{C2} & ~ & ~ & 3.31 & 5.81 & 7.09 & pure & 17 & ~ & 0.001 & 940 & 40 & ~ & ~ \\ 
        \textit{C3} & ~ & ~ & 3.41 & 5.85 & 7.14 & pure & 17 & ~ & 0.001 & 954 & 40 & ~ & ~ \\ 
        \textit{C4} & ~ & ~ & 3.37 & 6.01 & 7.34 & pure & 16 & ~ & 0.001 & 1076 & 40 & ~ & ~ \\ 
        \textit{C5} & ~ & ~ & 3.41 & 6.05 & 7.39 & pure & 16 & ~ & 0.001 & 393 & 40 & ~ & ~ \\ 
        \textit{C6} & ~ & ~ & 3.36 & 6.09 & 7.43 & mixed & 17 & ~ & 0.001 & 811 & 40 & ~ & ~ \\ 
        \textit{C7} & ~ & ~ & 3.25 & 6.05 & 7.39 & mixed & 17 & ~ & 0.001 & 729 & 40 & ~ & ~ \\ \hline
        \textbf{\textit{E1}} & S1 & FCC & 2.38 & 3.11 & 3.95 & - & - & 5.0 & 0.001 & 182 & 40 & MLC-1 & side wall touch \\ 
        \textbf{\textit{E2}} & ~ & ~ & 2.39 & 2.96 & 3.76 & - & - & 5.0 & 0.001 & 233 & 40 & ~ & ~ \\ 
        \textbf{\textit{E3}} & ~ & ~ & 2.38 & 3.15 & 4.00 & pure & 27 & 5, 500, 50 & SRJT & 322 & 40 & ~ & ~ \\ 
        \textit{E4} & ~ & ~ & 2.35 & 3.11 & 3.95 & mixed & 27 & 5, 500, 50 & SRJT & 429 & 400 & ~ & ~ \\ 
        \textbf{\textit{E5}} & ~ & ~ & 2.36 & 3.54 & 4.49 & pure & 25 & 5, 500, 50 & SRJT & 433 & 400 & ~ & ~ \\ \hline
        \textit{E6} & S1 & BCC & 2.45 & 3.64 & 4.62 & mixed & 24 & 5.0 & 0.001 & 1060 & 40 & MLC-1 & ~ \\ 
        \textbf{\textit{E7}} & ~ & ~ & 2.43 & 3.57 & 4.53 & pure & 24 & 5.0 & 0.001 & 2375 & 40 & ~ & ~ \\ 
        \textbf{\textit{E8}} & ~ & ~ & 2.41 & 3.59 & 4.56 & pure & 26 & 5, 500, 50 & SRJT & 1886 & 400 & ~ & ~ \\ 
        \textbf{\textit{E9}} & ~ & ~ & 2.39 & 3.49 & 4.43 & pure & 25 & 5, 500, 50 & SRJT & 1616 & 400 & ~ & ~ \\ 
        \textbf{\textit{E10}} & ~ & ~ & 2.38 & 3.36 & 4.26 & pure & 26 & 5, 500, 50 & SRJT & 1553 & 400 & ~ & ~ \\ \hline      
        \end{tabular} \caption{Pillar parameters. Avg.: average, TC: tilt corrected, $v$: velocity, Smp: sampling rate, SRJT: strain rate jump test, LO: lift-out. \textbf{Bold font} indicates the samples used for the SRS plot in Fig. \ref{fig:04}, \textit{italic font} represents the pillars selected for Fig. \ref{fig:03}.}\label{table:ta}
\end{table}

\begin{table}[!ht]
    \centering
    \begin{tabular}{|l|l|l|l|l|l|l|l|p{0.45in}|l|l|l|p{0.3in}|p{0.35in}|}
    \hline
        Pillar& ID & Phase & Avg. & Avg. & Height & Def. & Max. & Tip & SR & Yield & Smp & Load & Notes \\ 
        name & & & diam. & height & (TC)& mode & def.& $v$ & [/s] & stress & [Hz] & cell & \\ 
          & & & [$\mu$m] & [$\mu$m] & [$\mu$m] & & (\%) &  [nm/s] &  & [MPa] & & & \\ \hline
        \textit{F1} & S1 & BCC & 1.53 & 2.31 & 2.93 & mixed & - & 3.5 & 0.001 & 1268 & 40 & MLC-1 & side wall touch \\ 
        \textbf{\textit{F2}} & ~ & ~ & 1.53 & 2.32 & 2.94 & pure & 16 & 3.5 & 0.001 & 1987 & 40 & ~ & ~ \\ 
        \textbf{\textit{F3}} & ~ & ~ & 1.48 & 2.39 & 3.03 & pure & 16 & 350 & 0.1 & 2177 & 400 & ~ & ~ \\ 
        \textbf{\textit{F4}} & ~ & ~ & 1.48 & 2.06 & 2.61 & pure & 17 & 35 & 0.01 & 1988 & 400 & ~ & ~ \\ 
        \textbf{\textit{F5}} & ~ & ~ & 1.48 & 2.3 & 2.92 & pure & 16 & 350 & 0.1 & 2037 & 1000 & ~ & ~ \\ \hline
        \textbf{\textit{F6}} & S1 & FCC & 1.43 & 2.06 & 2.61 & pure & 18 & 3.5 & 0.001 & 359 & 40 & MLC-1 & ~ \\ 
        \textit{F7} & ~ & ~ & 1.41 & 2.2 & 2.79 & mixed & 17 & 3.5 & 0.001 & 528 & 40 & ~ & ~ \\ 
        \textbf{\textit{F8}} & ~ & ~ & 1.44 & 1.99 & 2.53 & - & - & 350 & 0.1 & 468.2 & 400 & ~ & ~ \\ 
        \textbf{\textit{F9}} & ~ & ~ & 1.42 & 1.87 & 2.37 & pure & 20 & 35 & 0.01 & 383 & 400 & ~ & ~ \\ 
        \textbf{\textit{F10}} & ~ & ~ & 1.43 & 1.6 & 2.03 & - & - & 350 & 0.1 & 443 & 1000 & ~ & ~ \\ \hline
        G1 & S1 & FCC & - & - & - & - & - & - & - & - & - & ~ & no data \\ 
        \textbf{G2} & ~ & ~ & 1.46 & 3.37 & 4.28 & pure & 14 & prop. & 12 & 866.7 & 100k & ST & ~ \\ 
        \textbf{G3} & ~ & ~ & 1.56 & 3.65 & 4.63 & pure & 12 & prop. & 6 & 520.1 & 50k & ~ & ~ \\ 
        \textbf{G4} & ~ & ~ & 1.53 & 3.37 & 4.28 & pure & 9 & prop. & 2 & 619.5 & 50k & ~ & ~ \\ 
        \textbf{G5} & ~ & ~ & 1.53 & 3.58 & 4.54 & pure & 30 & prop. & 2 & 489.8 & 50k & ~ & ~ \\ 
        \textbf{G6} & ~ & ~ & 2.55 & 3.19 & 4.05 & mixed & 21 & prop. & 16 & 561.8 & 100k & ~ & ~ \\ 
        \textbf{G7} & ~ & ~ & 2.93 & 3.97 & 5.04 & pure & 16 & prop. & 6 & 350.6 & 100k & ~ & ~ \\ 
        \textbf{G8} & ~ & ~ & 2.55 & 3.26 & 4.14 & pure & 21 & prop. & 2 & 479 & 50k & ~ & ~ \\ 
        \textbf{G9} & ~ & ~ & 2.53 & 3.24 & 4.11 & pure & 21 & prop. & 1 & 592 & 50k & ~ & ~ \\ 
        \textbf{G10} & ~ & ~ & 2.56 & 3.41 & 4.33 & mixed & 19 & prop. & 16 & 618.4 & 100k & ~ & ~ \\ \hline
        \textbf{H1} & S1 & BCC & 2.56 & 3.76 & 4.77 & mixed & 16 & prop. & 9 & 1220 & 100k & ST & ~ \\ 
        \textbf{H2} & ~ & ~ & 2.55 & 3.6 & 4.57 & mixed & 18 & prop. & 6 & 1180 & 100k & ~ & ~ \\ 
        \textbf{H3} & ~ & ~ & 2.63 & 4.04 & 5.13 & mixed & 17 & prop. & 2 & 897.8 & 50k & ~ & ~ \\ 
        \textbf{H4} & ~ & ~ & 2.63 & 4.05 & 5.14 & pure & 14 & prop. & 5 & 2274 & 50k & ~ & ~ \\ 
        \textbf{H5} & ~ & ~ & 2.7 & 4.15 & 5.27 & pure & 12 & prop. & 60 & 2064 & 100k & ~ & ~ \\ 
        \textbf{H6} & ~ & ~ & 1.48 & 2.36 & 2.99 & pure & 21 & prop. & 25 & 2742 & 100k & ~ & ~ \\ 
        \textbf{H7} & ~ & ~ & 1.46 & 1.62 & 2.06 & pure & 32 & prop. & 27 & 2557 & 100k & ~ & ~ \\ 
        \textbf{H8} & ~ & ~ & 1.59 & 2.49 & 3.16 & pure & 20/25 & prop. & 7 & 2182 & 50k & ~ & ~ \\ 
        \textbf{H9} & ~ & ~ & 1.58 & 2.64 & 3.35 & pure & 20/25 & prop. & 2 & 2191 & 50k & ~ & ~ \\ 
        \textbf{H10} & ~ & ~ & 1.61 & 2.64 & 3.35 & pure & 17 & prop. & 24 & 2085 & 100k & ~ & ~ \\ \hline
        I1 & S1 & FCC & 2.37 & 3.7 & 4.70 & - & - & 6$\times10^6$ & 1278 & - & 1M & ST & ringing \\ 
        I2 & ~ & ~ & 2.44 & 3.64 & 4.62 & - & - & 5.9$\times10^6$ & 1277 & - & 500k & ~ & ringing \\ 
        I3 & ~ & ~ & 2.48 & 3.91 & 4.96 & - & - & 5.9$\times10^6$ & 1189 & - & 500k & ~ & ringing \\ 
        \textbf{I4} & ~ & ~ & 2.46 & 3.59 & 4.56 & - & - & 6.0$\times10^6$ & 1317 & 732 & 500k & ~ & ~ \\ 
        \textbf{I5} & ~ & ~ & 2.47 & 3.28 & 4.16 & - & - & 6.0$\times10^6$ & 1442 & 607 & 500k & ~ & ~ \\ 
        I6 & ~ & ~ & 1.51 & 3.13 & 3.97 & - & - & 10$\times10^6$ & 2518 & 661 & 500k & ~ & ringing \\ 
        \textbf{I7} & ~ & ~ & 1.56 & 2.92 & 3.71 & - & - & 0.5$\times10^6$ & 135 & 475 & 50k & ~ & ~ \\ 
        I8 & ~ & ~ & 1.52 & 2.88 & 3.65 & - & - & 6.0$\times10^6$ & 1642 & - & 500k & ~ & ringing \\ 
        I9 & ~ & ~ & 1.56 & 3.75 & 4.76 & - & - & 6.0$\times10^6$ & 1261 & - & 500k & ~ & ringing \\ 
        I10 & ~ & ~ & 1.43 & 3.09 & 3.92 & - & - & 6.0$\times10^6$ & 1530 & - & 500k & ~ & ringing \\ \hline
    \end{tabular}\caption{Pillar parameters. Avg.: average, TC: tilt corrected, $v$: velocity, Smp: sampling rate, ST: SmarTip piezo based load cell, prop.: proportional loading profile. \textbf{Bold font} indicates the samples used for the SRS plot in Fig. \ref{fig:04}, \textit{italic font} represents the pillars selected for Fig. \ref{fig:03}.} \label{table:tb}
\end{table}

\begin{table}[!ht]
    \centering
    \begin{tabular}{|l|l|l|l|l|l|l|l|p{0.38in}|l|l|l|p{0.3in}|p{0.5in}|}
    \hline
        Pillar& ID & Phase & Avg. & Avg. & Height & Def. & Max. & Tip & SR & Yield & Smp & Load & Notes \\ 
        name & & & diam. & height & (TC)& mode & def.& $v$ & [/s] & stress & [Hz] & cell & \\ 
          & & & [$\mu$m] & [$\mu$m] & [$\mu$m] & & (\%) &  [mm/s] &  & [MPa] & & & \\ \hline
        \textbf{J1} & S1 & BCC & 2.44 & 4.05 & 5.14 & - & - & 6.0& 1167 & 2435 & 200k & ST & ~ \\ 
        \textbf{J2} & ~ & ~ & 2.48 & 4.32 & 5.48 & - & - & 6.0& 1094 & 2206 & 200k & ~ & ~ \\ 
        \textbf{J3} & ~ & ~ & 2.43 & 4.02 & 5.10 & - & - & 6.0& 1176 & 2575 & 500k & ~ & ~ \\ 
        \textbf{J4} & ~ & ~ & 2.39 & 3.99 & 5.06 & - & - & 6.0& 1185 & 1050 & 500k & ~ & ~ \\ 
        \textbf{J5} & ~ & ~ & 2.48 & 4.09 & 5.19 & - & - & 0.66& 127 & 1111 & 25k & ~ & ~ \\ 
        \textbf{J6} & ~ & ~ & 1.48 & 3.27 & 4.15 & - & - & 6.0& 1446 & 2859 & 500k & ~ & ~ \\ 
        J7 & ~ & ~ & 1.53 & 2.77 & 3.52 & - & - & 6.0& 1707 & - & 500k & ~ & ~ \\ 
        J8-J10 & ~ & ~ & - & - & - & - & - & - & - & ~ & - & ~ & destroyed \\ 
        \textbf{J11} & ~ & ~ & 1.55 & 2.65 & 3.36 & - & - & 6.0 & 1784 & 2642 & 500k & ~ & ~ \\ \hline
        \textbf{K1} & S1 & BCC & 2.56 & 2.81 & 3.43 & - & - & 30 & 8746 & 3961 & 1M & ST & ~ \\ 
        \textbf{K2} & ~ & ~ & 2.46 & 2.57 & 3.14 & - & - & 30 & 9562 & 4740 & 1M & ST & ~ \\ 
        \textbf{K3} & ~ & ~ & 2.52 & 2.75 & 3.36 & pure & - & 30 & 0.01 & 1793 & 40 & SLC & LO/TEM \\ 
        K4 & ~ & ~ & 2.41 & 2.59 & 3.16 & - & - & 30 & 9488 & - & 1M & ST & no def., LO/TEM \\ 
        \textit{K5} & ~ & ~ & 2.38 & 2 & 2.44 & mixed & - & 3$\times10^{-5}$ & 0.01 & 848.2 & 40 & SLC & ~ \\ 
        \textbf{K6} & ~ & ~ & 2.4 & 2.49 & 3.04 & - & - & 3$\times10^{-5}$ & 0.01 & 2214 & 40 & SLC & LO/TEM \\
        \textbf{K7} & ~ & \textcolor{gray}{FCC?} & 2.35 & 2.29 & 2.80 & mixed & - & 10 & 3577 & 661 & 500k & ST & pancake\\ 
        \textbf{K8} & ~ & ~ & 1.63 & 2.56 & 3.13 & - & - & 10 & 3200 & 2618 & 500k & ST & pancake \\ 
        \textbf{K9} & ~ & ~ & 1.58 & 2.38 & 2.91 & - & - & 10 & 3442 & 2740 & 500k & ST & pancake \\ 
        \textbf{K10} & ~ & ~ & 1.59 & 2.34 & 2.86 & - & - & 10 & 3501 & 3033 & 500k & ST & pancake \\ 
        \textbf{K11} & ~ & \textcolor{gray}{FCC?} & 1.54 & 2.11 & 2.58 & mixed & - & 30 & 11647 & 1420 & 1M & EST & ~ \\ 
        \textbf{K12} & ~ & ~ & 1.58 & 2.53 & 3.09 & - & - & 30 & 9713 & 5600 & 1M & EST & ~ \\ 
        \textbf{K13} & ~ & ~ & 1.53 & 2.35 & 2.87 & - & - & 50 & 17429 & 6700 & 1M & EST & ~ \\ 
        \textbf{K14} & ~ & ~ & 1.52 & 1.98 & 2.42 & - & - & 10 & 4137 & 2580 & 1M & EST & ~ \\ 
        \textbf{K15} & ~ & ~ & 1.52 & 2.1 & 2.56 & - & - & 10 & 3901 & 3330 & 1M & EST & LO/TEM \\ \hline
        \textbf{M1} & S1 & FCC & 1.47 & 3.11 & 3.95 & - & - & 10 & 2534 & 918 & 1M & ST & ~ \\ 
        \textbf{M2} & ~ & ~ & 1.43 & 3.14 & 3.98 & - & - & 10 & 2510 & 822 & ~ & ~ & ~ \\ 
        \textbf{M3} & ~ & ~ & 1.45 & 3.54 & 4.49 & - & - & 30 & 6678 & 1260 & ~ & ~ & ~ \\ 
        \textbf{M4} & ~ & ~ & 1.49 & 3.34 & 4.24 & - & - & 30 & 7078 & 1169 & ~ & ~ & ~ \\ 
        M5 & ~ & ~ & 1.52 & 3.58 & 4.54 & - & - & 50 & 11006 & - & ~ & ~ & destroyed \\ 
        \textbf{M6} & ~ & ~ & 1.46 & 3.49 & 4.43 & - & - & 50 & 11290 & 1074 & ~ & ~ & ~ \\ 
        \textbf{M7} & ~ & ~ & 1.49 & 3.4 & 4.31 & - & - & 50 & 11589 & 1180 & ~ & ~ & ~ \\ 
        \textbf{M8} & ~ & ~ & 1.44 & 2.9 & 3.68 & - & - & 30 & 8152 & 1250 & ~ & ~ & ~ \\ 
        \textbf{M9} & ~ & ~ & 1.46 & 3.33 & 4.23 & - & - & 50 & 11832 & 1106 & ~ & ~ & ~ \\ 
        \textbf{M10} & ~ & ~ & 1.43 & 3.37 & 4.28 & - & - & 50 & 11692 & 1000 & ~ & ~ & ~ \\  \hline
    \end{tabular}\caption{Pillar parameters. Avg.: average, TC: tilt corrected, $v$: velocity, Smp: sampling rate, ST: SmarTip piezo based load cell, EST: Enhanced SmarTip, LO: lift-out. \textbf{Bold font} indicates the samples used for the SRS plot in Fig. \ref{fig:04}.} \label{table:tc}
\end{table}

\end{document}